\documentclass{article}

\usepackage{RR}
\usepackage{graphicx,color,natbib,amssymb,subfigure,amsmath,bm}
\usepackage[french]{babel}
\usepackage{hyperref}
\usepackage{times}

\renewcommand{\(}{\left (}
\renewcommand{\)}{\right )}

\newcommand{\GAL}[1]{\bm{{#1}^{\scriptscriptstyle G}}}
\newcommand{\CAL}[1]{\bm{{#1}^{\scriptscriptstyle C}}}
\newcommand{\ERR}[1]{\bm{{#1}^{\scriptscriptstyle E}}}
\newcommand{\UNK}[1]{\bm{{#1}}}
\newcommand{\AG}{\GAL{A}}
\newcommand{\BG}{\GAL{B}}
\newcommand{\CG}{\GAL{C}}
\newcommand{\HG}{\GAL{H}}
\newcommand{\EG}{\GAL{E}}
\newcommand{\FG}{\GAL{F}}
\newcommand{\GG}{\GAL{G}}
\newcommand{\XG}{\GAL{X}}
\newcommand{\AC}{\CAL{A}}
\newcommand{\CC}{\CAL{C}}

\newcommand{\GC}{\CAL{G}}
\newcommand{\XC}{\CAL{X}}
\newcommand{\BU}{\UNK{B}}
\newcommand{\XU}{\UNK{X}}
\renewcommand{\AE}{\ERR{A}}
\newcommand{\CE}{\ERR{C}}
\newcommand{\GE}{\ERR{G}}
\newcommand{\pl}{\partial}
\newcommand{\h}[1]{\hat{#1}}
\newcommand{\ab}{\bm{a}}
\newcommand{\ub}{\bm{u}}
\newcommand{\ubb}{\bm{\bar{u}}}
\newcommand{\vb}{\bm{v}}
\newcommand{\vbb}{\bm{\bar{v}}}
\newcommand{\fb}{\bm{f}}
\newcommand{\ah}{\h{a}}
\newcommand{\ch}{\h{c}}
\newcommand{\ahb}{\bm{\ah}}
\newcommand{\XB}{\bm{X}}
\newcommand{\phib}{\bm{\Phi}}
\newcommand{\xb}{\bm{x}}
\newcommand{\wb}{\bm{w}}
\newcommand{\nb}{\bm{n}}
\newcommand{\Kb}{\mathsf{K}}
\newcommand{\um}{\overline{u} }
\newcommand{\uc}{u_c }
\newcommand{\phis}{{\phib}^s }
\newcommand{\phik}{{\phib}^k }
\newcommand{\phir}{{\phib}^r }
\newcommand{\ii}{\textsc{i}}
\newcommand{\fbac}{\fb(\ab(t),c(t),\dd{c}(t))}
\newcommand{\fbahc}{\fb(\ahb(t),c(t),\dd{c}(t))}
\newcommand{\fbahch}{\fb(\ahb(t),\ch(t),\dd{\ch}(t))}
\newcommand{\fbaci}{\fb(\ahb(t),c^{\ii}(t),\dd{c}^{\ii}(t))}
\newcommand{\ff}{\textsc{f}}
\newcommand{\fbacf}{\fb(\ab(t),c^{\ff}(t),\dd{c^{\ff}}(t))}
\newcommand{\M}[1]{\mathcal{#1}}
\newcommand{\grad}{\nabla}
\newcommand{\lap}{\Delta}
\newcommand{\re}{\frac{1}{Re}}
\renewcommand{\d}[1]{\, d #1}
\newcommand{\dd}[1]{\dot{#1}}
\renewcommand{\and}{\mbox{ and }}
\newcommand{\for}{\mbox{ for }}
\newcommand{\on}{\mbox{ on }}
\newcommand{\st}{\mbox{ subject to }}
\renewcommand{\c}{\cdot}
\newcommand{\mkbox}[3]{
  \newsavebox{#1}
  \sbox{#1}{\raisebox{0cm}[#2][#3]{ }}
}
\mkbox{\mybox}{0.5cm}{0.3cm}
\newcommand{\lspace}{\usebox{\mybox}}  

\newenvironment{syst}{\begin{eqnarray}
    \begin{gathered} \displaystyle }{\end{gathered} \end{eqnarray} }
\newenvironment{systa}{\begin{eqnarray} 
    \begin{gathered}  \begin{array}{rcl} \displaystyle}{\end{array} \end{gathered}
\end{eqnarray} }
\newenvironment{systa*}{\begin{eqnarray*} 
    \begin{gathered}  \begin{array}{rcl} \displaystyle}{\end{array} \end{gathered}
\end{eqnarray*} }

\RRdate{Juin 2008}
\RRauthor{Jessie Weller\thanks[imbinria]{IMB - Universit\'{e} Bordeaux et MC2
    - INRIA Bordeaux Sud-Ouest, 33405 Talence, France} 
  \and
  Edoardo Lombardi\thanksref{imbinria}
  \and
  Angelo Iollo\thanksref{imbinria}
}
\authorhead{Weller \& Lombardi \& Iollo}
\RRtitle{Identification robuste de sillages tourbillonnaires controll\'es} 
\RRetitle{Robust model identification of actuated vortex wakes}
\titlehead{Robust model identification of actuated vortex wakes}
\RRresume{Nous proposons une m\'ethode de r\'eduction de mod\`ele,
  pour des \'ecoulements actionn\'es, bas\'ee sur la r\'egularisation
  d'un probl\`eme inverse. Le probl\`eme inverse vise \`a minimiser
  l'erreur entre les pr\'edictions du mod\`ele et des simulations de
  r\'ef\'erence. Les param\`etres \`a identifier sont les coefficients
  d'une expansion polynomiale qui mod\'elise les dynamiques
  temporelles d'un nombre limit\'e de modes globales. Ces modes sont
  obtenus par D\'ecomposition Orthogonale aux valeurs Propres
  (POD). Il s'agit d'une m\'ethode pour calculer les \'el\'ements les
  plus repr\'esentatifs, en termes d'\'energie, d'une base de
  donn\'ees de simulations. Il est montr\'e que des mod\`eles bas\'es
  sur une simple projection de Galerkin et sur des techniques
  classiques de calibration ne sont pas viables. Ils sont soit mal
  pos\'es, soit donnent une approximation erron\'ee de la solution d\`es
  qu'ils sont utilis\'es pour pr\'edire des dynamiques n'appartenant
  pas \`a la base de donn\'ees originale. En revanche, des \'evidences
  num\'eriques montrent que la m\'ethode que nous proposons permet de
  construire des mod\`eles robustes, qui r\'eagissent correctement \`a des
  variations de param\`etres, et pourraient donc être utilis\'es pour des
  probl\`emes de contr\^ole d'\'ecoulement.} 
\RRabstract{We present a low-order modeling technique for actuated flows based
  on the regularization of an inverse problem. The inverse problem
  aims at minimizing the error between the model predictions and
  some reference simulations. The parameters to be identified are a
  subset of the coefficients of a polynomial expansion which models
  the temporal dynamics of a small number of global modes. These
  global modes are found by Proper Orthogonal Decomposition, which
  is a method to compute the most representative elements of an
  existing simulation database in terms of energy. It is shown that
  low-order control models based on a simple Galerkin projection and
  usual calibration techniques are not viable. They are either
  ill-posed or they give a poor approximation of the solution as
  soon as they are used to predict cases not belonging to the
  original solution database. In contrast, numerical evidence shows
  that the method we propose is robust with respect to variations of
  the control laws applied, thus allowing the actual use of such
  models for control.}
\RRmotcle{mod\`eles r\'eduits, contr\^ole, probl\`emes inverses}  
\RRkeyword{reduced order models, control, inverse problems}
\RRprojet{MC2}
\RRtheme{\THNum} % cas d'un seul theme
\RCBordeaux % pour ceux qui sont dans le virtuel
\begin{document}
\makeRR   % cas d'un rapport de recherche
%-----------------------------------------------------------------------------
\section{Introduction}
\label{sec:intro}

We consider the problem of describing the dynamics of an infinite
dimensional system using a small number of degrees of freedom. In
particular we concentrate on the problem of devising accurate and
robust models of actuated fluid flows past  bluff obstacles. These
flows are dominated by the presence of large-scale vortices due to
massive separation, and are good candidates for a low-dimensional
representation. The point of view that we privilege is empiric: the
functional space in which we seek the low-dimensional solution is
derived using proper orthogonal decomposition (POD) \cite{L67}. POD
makes use of simulation databases to determine optimal functional
spaces in terms of solution representation. A vast literature
concerning this way of modeling fluid flows exists
\cite{Galletti2006,Buffoni2006,Galletti2004,MK02}, and some results
show the possible interest of using POD in applications such as flow
control \cite{B06,G98,GR98}. \\
However, several problems related to the idea of modeling a flow by a
small number of variables are open. One of the issues is the
asymptotic stability of the models obtained. Often such models are
capable of correctly reproducing the dynamics over small time
intervals, whereas the asymptotic behavior converges to incorrect
limit cycles~\cite{MK02}. This issue is related to both numerical
artifacts and to an improper representation of the
solution~\cite{Iollo2000,rempfer2000,Noack}. As a results low-order models
are of delicate use and not robust to parameter variations. \\
The present study describes a method to obtain robust low-order
models.  In previous works we showed that it is possible to obtain
accurate low-order models of relatively complicated flows by
minimizing the error between the model results and the reference
solution \cite{Buffoni2006}. Here we extend those works to cases where
the flow is actuated by devices that can affect locally or globally the
velocity and pressure fields. The objective is to derive a low-order
model that provides accurate predictions and that is robust to
variations of the control law employed. The main idea is to identify
the manifold over which the non-linear dynamics of the POD modes lies,
when the input to the system is varied. In this spirit, several
dynamics are included in the identification procedure coupled with a
Tikhonov type regularization. The case of a precomputed control as
well as the case of a feed-back control are studied. \\
The practical relevance of this work is that low-order  models make
possible to devise or to optimize controls for large-scale problems
that would not be otherwise solvable in terms of computational
size. Applications of this method is straight forward for models other
than the Navier-Stokes equations. 

\section{Reduced Order Modeling using POD}
\label{sect:gal}

\subsection{Flow setup}
\label{sec:flow}

\begin{figure}[!t]
  \centering
  \vspace{-4.cm}
  \includegraphics[width=8.cm]{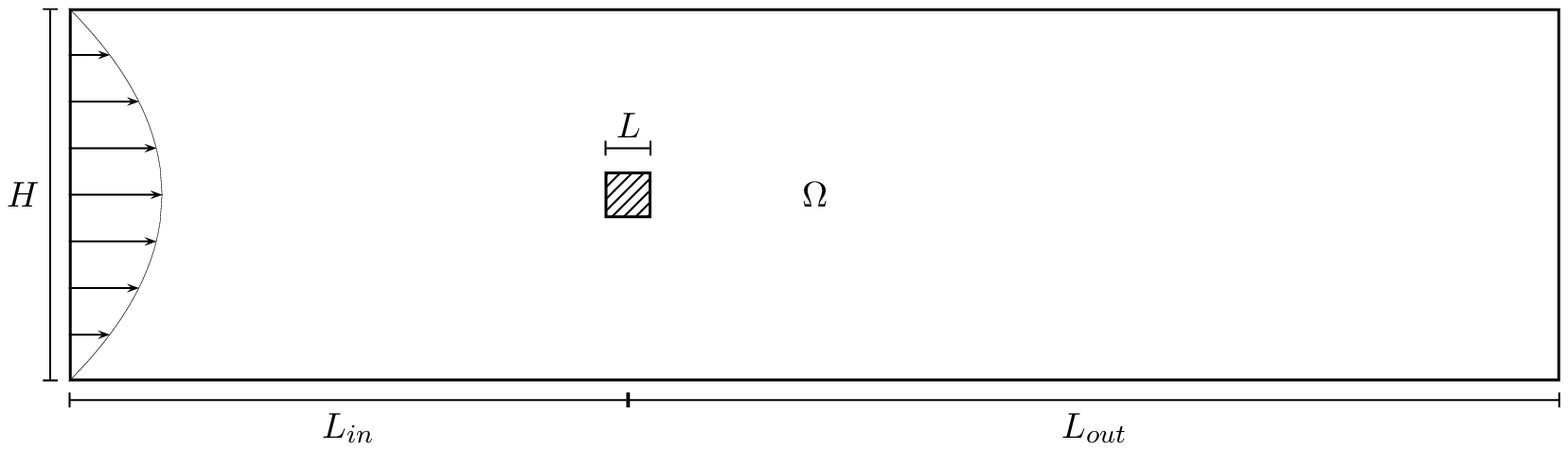}
  \vspace{-5.cm}
  \caption{Computational domain $\Omega$.}
  \label{fig:setup} 
\end{figure}

\begin{figure}[!t]
  \begin{center}
    \vspace{-6.cm}
    \includegraphics[clip, height=16.cm]{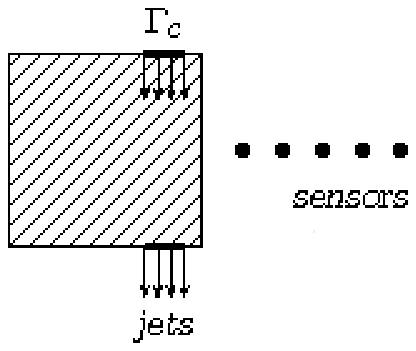}
    \vspace{-7.cm}
    \caption{Placement of synthetic jet and sensors for
      control} 
    \label{fig:jets}
  \end{center}
\end{figure}

We consider a two-dimensional laminar flow past a confined square cylinder. This
setup presents a reasonable compromise between physical complexity and
computational cost. A sketch showing the geometry, the frame of
reference and the adopted notation is plotted in Fig.\ref{fig:setup}.  
At the inlet, the incoming flow is assumed to have a Poiseuille
profile with maximum center-line velocity $U$. With reference to
Fig.\ref{fig:setup}, $L/H=1/8$, $L_{in}/L = 12$, $L_{out}/L =
20$. No-slip conditions are enforced both on the cylinder and on the
parallel walls. Details concerning the grids and the numerical set up
are reported in \cite{Buffoni2006}.\\
All the quantities mentioned in the following have been made
non-dimensional by $L$ and $U$. The two-dimensional unforced flow
obtained is a classic vortex street with a well defined shedding
frequency. The interaction with the confining walls leads to some
peculiar features, like the fact that the vertical position of the
span-wise vortices is opposite to the one in the classic von
K\'arm\'an street \cite{Camarri2006}.\\
The presence of an actuator is modeled by imposing a new boundary
condition on a small surface $\Gamma_c$ of $\pl{\Omega}$:
\[ \ub(\xb,t) \cdot \nb(\xb) = c(t), \quad \xb \in \Gamma_c \]
For control purposes we place two actuators on the cylinder. They are
driven in opposite phase, as shown in Fig. \ref{fig:jets}: 
\[ v(\xb,t) = c (t) , \quad \xb \in \Gamma_c  \] 
The control law $c(t)$ can be precomputed, or obtained using a
proportional feedback law. For example, using measurements of the vertical velocity
at points $\xb_j$ in the cylinder wake, we can define a proportional
control law:   
\[ c(t) = \sum_{j=1}^{N_s} \Kb_j v(\xb_j,t) \] 
where $N_s$ denotes the number of sensors used.

\subsection{Proper Orthogonal decomposition with the snapshot method}

Seeking a reduced order solution that can be written
\[ \ub_R(\xb,t) = \sum_{r=1}^{N_r} a_r (t) \phib^r (\xb) \]
where the spatial functions $\phib^r$ are obtained by POD has become a
popular approach when dealing with large scale systems. A vast
literature concerning the POD procedure exists \cite{Holmes}, we
refer to these works for a more general review of the method.
%\cite{L67,AHLS88}
\subsubsection{The POD basis}
\label{sub:podbasis}

In our case, a numerical simulation of the Navier-Stokes equations is
performed over a time interval $[0,T]$, and the velocity field is
saved at $N_t$ time instants $t_i \in [0,T]$. This yields a data set
$\{\ub^i(\xb) = \ub(\xb,t^i) \}_{i=1..N_t}$. The aim of the POD
procedure is to find a low dimensional subspace of $\M{L} =
span\{\ub^1, \cdots, \ub^{N_t}\}$, that gives the best approximation
of $\M{L}$. We therefore seek an orthonormal set
$\{\phib^r\}_{r=1...N_r}$, where $N_r << N_t$, and a set of
coefficients $\ah_k^i$ such that the reconstruction error: 
\begin{equation}
  \label{rerr} 
  \sum_{i=1}^{N_t} {\left \| \ub^i - \sum_{r=1}^{N_r} \ah_k^i \phib^r
    \right \|}^2_{L^2(\Omega)} 
\end{equation}
is minimal.\\
Following Sirovich's idea  \cite{S87} the POD modes are expressed as
linear combinations of the snapshots: 
 \[ \phib^k(\xb) = \sum_{i=1}^{N_t} b_i^k \ub^i(\xb) \for k=1..N_r\]
The vectors ${[b^r_i]}_{i=1\cdots N_t}$ are found to be the
eigenvectors of the $N_t \times N_t$ correlation matrix $R$, $R_{ij} =
(\ub^i,\ub^j)$, corresponding to the $N_r$ highest eigenvalues, while
the $\ah_k^i$ are equal to the scalar products $(\ub^i,\phib^r)$.\\
In the case of forced flow, the snapshots depend on the control law
$c(t)$ used. In this work we consider POD basis derived from numerical
simulations obtained using several different control laws. Indeed,
there are a number of other parameters that could be varied, but since
our aim is to study the effect of a control law, we set ourselves in
the following framework: 
\renewcommand{\labelitemi}{\textendash}
\begin{itemize}
\item Time instants, Reynolds number, domain geometry, placement of the
  actuators will be the same for all the snapshots in the database.
\item The control law $c(t)$ will be varied 
\end{itemize}
The data set used for the POD is therefore written: 
\begin{equation*}
  \{\ub^{i,\ell}(\xb) =\}_{i=1..N_t,\ell=1..N_c}
\end{equation*}
where $N_c$ denotes the number of control laws considered. If
$\M{C}=\{c_1,c_2,\cdots,c_{N_{c}}\}$ is the set of control laws used
to obtain the database, the ensuing POD basis is denoted $\phib
({\M{C}})$. In the first part of this work, $\M{C}$ is reduced to a
single element which we denote $c(t)$.

\subsubsection{Dealing with the boundary conditions}

In the non-controlled case, we lift the boundary conditions on the
velocity fields by defining a new set of snapshots:
\[ \wb^i(\xb) = \ub^i(\xb) - \ubb(\xb) \]
where $\ubb$ is some reference velocity field that satisfies the same
boundary conditions as the snapshots. In the present configuration, it
can be the steady unstable solution, or a time average of the
snapshots $\ub^k$.\\ 
When an extra boundary condition is imposed on the cylinder for
control purposes, the snapshots are chosen to be: 
\[ \wb^i(\xb) = \ub^i(\xb) - \ubb(\xb) - c(t^i) \ub_c(\xb) \]
where $\ub_c(\xb)$ satisfies the following criteria:
\[ \ub_c(\xb) = \ubb(\xb) \on \Gamma \backslash \Gamma_c  \and
\ub_c(\xb) = 1 \on \Gamma_c  \] 
In practice we use the velocity field proposed in \cite{Galletti2006}:
\[ \ub_c(\xb) = \frac{1}{c^{\star}} (\ubb^{\prime}(\xb) - \ubb(\xb)) \]
where $\ubb^{\prime}$ is obtained in the same way as $\ubb$ but
applying a constant control equal to $c^{\star}$ on $\Gamma_c$ . The
low-dimensional solution is now written: 
\begin{equation}
  \label{gal_rap}
  \ub_R(\xb,t) = \ubb(\xb) + c(t) \ub_c(\xb) + \sum_{k=1}^{N_r} a_k (t) \phib^k (\xb)
\end{equation}

\subsection{POD-Galerkin Reduced Order Model}

Galerkin projection of the incompressible Navier-Stokes equations onto
the first $N_r$ POD modes yields a system of ordinary differential equations:
\begin{eqnarray} 
  \label{eq:gal_rom}
  \begin{gathered}  
    \left \{ 
    \begin{array}{@{\lspace}rcl}
      \displaystyle 
      \dd{a}_r (t) &=& \AG_r + \CG_{kr} a_k(t) + \BG_{ksr} a_k(t) a_s(t) +
      \M{P}_r \\
      & & + \quad \EG_r \dd{c}(t)  + \FG_r c^2(t) + \GG_r c(t)  + \HG_{kr}
      a_k(t) c(t) \\
      a_r(0) &=& a_r^0 \\
      \multicolumn{3}{@{\lspace}l}{1 \leq r \leq N_r}
    \end{array}
    \right .
  \end{gathered}
\end{eqnarray}
where:
\begin{eqnarray*}
  & & \AG_r = - ((\um \c \grad) \um , \phir ) + \re (\lap \um,\phir) \\
  & & \BG_{ksr} = - ((\phik \c \grad) \phis , \phir )\\
  & & \CG_{kr} = - ((\um \c \grad) \phik , \phir ) - ((\phik \c \grad) \um ,
  \phir )  + \re (\lap \phik,\phir) \\
  & & \EG_r = (\uc, \phir)\\
  & & \GG_r = - ((\um \c \grad) \uc , \phir ) - ((\uc \c \grad) \um ,
  \phir ) + \re (\lap \uc,\phir) \\
  & & \FG_r = ((\uc \c \grad) \uc , \phir ) \\
  & & \HG_{rk} = ((\uc \c \grad) \phik , \phir ) + ((\phik \c \grad) \uc ,
  \phir ) \\
  & & \M{P}_r = (\nabla p, \phib^r)
\end{eqnarray*}
We note that since the snapshots satisfy the continuity equation, the
modes do also. This implies that the pressure term $\M{P}_r$ is equal
to $\int_{\pl \Omega} p \phib^r \d{s}$. If velocity field is constant
at the boundaries, the POD modes are zero there. The pressure term
therefore disappears completely.\\
Setting:
\begin{eqnarray*}
  \XG_r &=& {\left [ \AG_r ,\, {\{\BG_{ksr}\}}_{k,s=1\cdots N_r},  \, {\{\CG_{k
          r}\}}_{k=1\cdots N_r}, \,\EG_r, \, \FG_r, \, \GG_r , \, {\{\HG_{k
          r}\}}_{k=1\cdots N_r} \right ] }^t
\end{eqnarray*}
and
\begin{eqnarray*}
  \fbac &=& \left [ 1, \, {\{a_k(t)a_s(t) \}}_{k,s=1\cdots N_r}, \,
    {\{a_k (t)\}}_{k=1\cdots N_r}, \right .\\
    & & \left . \quad  \quad \quad  \dd{c}(t), \, c^2(t), \, c(t),  \, {\{a_k(t)
      c(t)\}}_{k=1\cdots N_r} \right ]
\end{eqnarray*}
the first equation in  (\ref{eq:gal_rom}) can be written in the
compact form:
\[ \dd{a}_r (t) = \fbac \cdot \XG_r \]
The initial value problem (\ref{eq:gal_rom}) is a reduced order model
of the Navier-Stokes equations, called the POD-Galerkin model. Such a
model might be inaccurate for it may not take into account enough of
the dynamics. Indeed, although a number $N_r$ of modes can be
sufficient to capture most of the flow energy, the neglected modes
continue to play an important role in the flow dynamics through their
interaction with the conserved ones. The difference between the
solutions $a_r(t)$ of (\ref{eq:gal_rom}), and the coefficients
$\ah_r(t)$ obtained by projecting the numerical data onto the POD
modes ($\ah_r(t^i) = \ah_r^i$), has been underlined in several papers
\cite{Galletti2006,B06,Couplet2005}. It is therefore interesting to
build a model that exploits the knowledge one has of the dynamics,
that is the set of temporal projection coefficients $\ah_r^i$. This is
the subject of the next section. 

\section{Robust low order models}

\subsection{Calibration method}

The idea of calibration is to keep the structure of 
the above model while adjusting the coefficients
of the system so its solution is closer to the desired one. In
previous work \cite{Galletti2004,Galletti2006}, it was shown that
robust low order models could be obtained by solving the minimization
problem: 
\begin{syst}
  \label{eq:adj_cal}
  \min_{a,\XU} \sum_{r=1}^{N_r} \int_0^T {\(a_r(t)-\ah_r(t) \)}^2 \d{t}
  \\
  \st  \qquad \dd{a}_r (t) =\fbac \cdot \XU_r 
\end{syst}
This \textsl{state calibration} method, which involves solving a
strongly non-linear system, works well as long as the number of
snapshots considered remains limited. For a large number of snapshots,
the computational costs are excessive. \\
Another method was suggested in \cite{Galletti2004}, and has been
experimented, for a case with no control, in
\cite{Buffoni2006,Buffoni2008,Couplet2005} with good results. It
consists in choosing $\XU$ as the solution of: 
\begin{eqnarray}
  \label{eq:adot}
  \min_{\XU} \sum_{r=1}^{N_r}  \int_0^T {\(\dd{\ah}_r(t)-
    \fbahc \cdot \XU_r \)}^2 \d{t} 
\end{eqnarray}
This method can be interpreted as approximating the error 
\[ e_r (t) =  \dd{\ah}_r(t)-\fbahc \cdot \XG_r\]
by a quadratic function of all the non-discarded temporal
coefficients, $c(t)$ and $\dd{c}(t)$. Other choices for the
approximation of $e_r$ lead to partial calibration problems.\\  
For example, if we suppose $e_r \approx \AE_r + \CE_{k r}  a_k + \GE c
$ then we will solve:
\begin{eqnarray}
\label{eq:padot}
\displaystyle \min_{\XC_1} \sum_{r=1}^{N_r}  \int_0^T {\left (\dd{\ah}_r(t)  -
    \fb_1(t) \cdot \XC_{r,1} - \fb_2(t) \cdot \XG_{r,2} \right )}^2 \d{t} 
\end{eqnarray}
where
\begin{systa*}
  \XC_{r,1} &=& {\left [ \AC_r, \, {\{\CC_{k r}\}}_{k=1\cdots N_r},
      \,  \GC_r \right ]}^t \\
  \XG_{r,2} &=& {\left [{\{\BG_{ks
          r}\}}_{k,s=1\cdots N_r},\,  \EG_r, \, \FG_r, \, {\{\HG_{k r}\}}_{k=1\cdots N_r}  \right]}^t
\end{systa*}
and
\begin{systa*}
  \fb^1(t) &=& \left [ 1, \, {\{a_k (t)\}}_{k=1\cdots N_r}, \,  c(t)  \right
  ] \\
  \fb^2(t) &=& \left [{\{a_k(t) a_s(t) \}}_{k,s=1\cdots N_r} ,
    \,  \dd{c}(t) , \,  c²(t) , \, {\{a_k (t) c(t)\}}_{k=1\cdots N_r}  \right ]
\end{systa*}

Of course, other choices of which terms to calibrate or not can be
made. For a general formulation we denote $N_{cal}$ the number of
terms of vector $\XU_r$ that are calibrated, and we have $N_{cal}
\leqslant N_r^2 + 2 \times N_r + 4$. Whatever the choice for
$N_{cal}$, this approach is always much more efficient than
(\ref{eq:adj_cal}) since it involves solving $N_r$ linear symmetric
systems of size $N_{cal}^2$: 
\begin{syst}
  \label{eq:calsolve}
  \int_0^T \fb_1^t(t) \fb_1(t) \d{t} \quad \XC_{r,1} = \int_0^T \fb_1^t(t) \(\dd{\ah}_r(t) - \fb_2(t)
  \cdot \XG_{r,2} \)  \d{t} 
\end{syst}
 The more terms of $\XB_r$ are calibrated, the more the problem
 becomes ill-conditioned. For this reason we choose not to calibrate
 the  $N_r^3$ terms $\BU_{ksr}$.\\

Once the model has been calibrated to fit a particular control law
$c(t)$, it can of course be integrated using another control law. Denoting the input control law $c^{\ii}(t)$,
the calibrated model is written:
\begin{eqnarray} 
  \label{eq:rom}
  \begin{gathered} 
    \M{R}({\{c\}})
    \left \{ 
    \begin{array}{@{\lspace}rcl}
      \displaystyle 
      \dd{a}_r (t) &=& \fbaci \cdot \XU_r  \\
      a_r(0) &=& a_r^0 \\
      \multicolumn{3}{@{\lspace}l}{1 \leq r \leq N_r}
    \end{array}
    \right .
  \end{gathered}
\end{eqnarray}
where by denoting $\M{R}({\{c\}})$ the model we put in evidence that
it was calibrated using the control $c(t)$.

\subsection{Well-posedness and robustness}
\label{sub:cal_2}

\subsubsection{Calibration with feedback control laws}
\label{sub:cal_fbck}
We suppose that the control is obtained using a proportional feedback
law (Sec.\ref{sec:flow}): 
\[ c(t) = \sum_{j=1}^{N_s} \Kb_j v(\xb_j,t) \]  
We can now consider two different calibration problems. The first is
the problem (\ref{eq:adot}), the second is: 
\begin{eqnarray}
  \label{eq:fbck_adot}
  \min_{\XU} \sum_{r=1}^{N_r}  \int_0^T {\(\dd{\ah}_r(t)-
    \fbahch \cdot \XU_r \)}^2 \d{t} 
\end{eqnarray}
where $\ch$ is defined by:
\begin{equation}
  \label{fbc}
  \ch(t) =  \Kb_j v_R(\xb_j,t) = \Kb_j \(\vbb(\xb_j) + \ch(t) \vb_c(\xb_j) + \sum_{r=1}^{N_r}
  \ah_r(t) \phib^r_v(\xb_j)\) 
\end{equation}
This last approach makes the reduced order model a feedback model,
which is useful if we want to use the model to determine an optimal
feedback law. The problem is however under-determined.\\
We reformulate (\ref{fbc}) to clearly show the dependency of $\ch$ on
$\ahb$: 
\begin{equation}
  \label{eq:fbcc}
  \ch(t) =  \kappa_0 + \sum_{r=1}^{N_r} \kappa_r \ah_r(t)
\end{equation}
where
\[ \kappa_0 = \frac{\Kb_j}{1 - \Kb_i \vb_c(\xb_i)} \quad \vbb(\xb_j)
\quad \and \quad  \kappa_r  = \frac{\Kb_j}{1 - \Kb_i
  \vb_c(\xb_i)} \quad \phib^r_v(\xb_j) \]
We now look at the partial-calibration problem described above.
The function $\fb^1$ that appears in
system (\ref{eq:calsolve}) can be reformulated:
\[ \fb^1(t) = \left [ 1, \, {\{\ahb_k (t)\}}_{k=1\cdots
    N_r}, \,  \kappa_0 + \kappa_{\ell} \ahb_{\ell}(t)  \right
] \]
System (\ref{eq:calsolve}) is therefore rank deficient. The problem
remains if more of the system coefficients are calibrated, and
according to the choice made the rank of the problem matrix can even
diminish with respect to $N_{cal}$.\\ 
This difficulty can however be solved by using one of the two methods
proposed in the section (\ref{sub:cal_3}). Finally, the proportional
feedback reduced order model is written:
\begin{eqnarray} 
  \label{eq:romf}
  \begin{gathered} 
    \M{R}^{\ff}({\{\ch\}})
    \left \{ 
    \begin{array}{@{\lspace}rcl}
      \displaystyle 
      \dd{a}_r (t) &=& \fbacf \cdot \XU_r  \\
      c^{\ff}(t) &=& \displaystyle \sum_{j=1}^{N_s} \Kb_j \(\vbb(\xb_j) + c^{\ff}(t) \vb_c(\xb_j) + \sum_{r=1}^{N_r}
      a_r(t) \phib^r_v(\xb_j)\) \\
      a_r(0) &=& a_r^0 , \quad c^{\ff}(0) = \ch(0)\\
      \multicolumn{3}{@{\lspace}l}{1 \leq r \leq N_r}
    \end{array}
    \right .
  \end{gathered}
\end{eqnarray}

\subsubsection{Instability issues}

The system solved for calibration can be ill-posed even in cases
different to the one just described. To understand why this is, it is
sufficient to go back to the \textsl{state calibration method}
mentioned at the beginning of the section. Solving the minimization
problem (\ref{eq:adj_cal}) involves solving a non-linear system for
which the uniqueness of solution is not guaranteed. The state
calibration functional can therefore have several local optima, and so
there are several possible choices for $\XU$ that will lead to a low
value of the error $\|\ahb-\ab\|$. Since these choices should also be
good choices for the minimization problem (\ref{eq:padot}), the
matrix $\int_0^T \fb_1^t \fb_1 \d{t}$ in (\ref{eq:calsolve}) is
in general almost singular. A model obtained by inverting this matrix
is most often  very unstable. To overcome this problem we propose a
Tikhonov type regularization method which we describe in the next
section. 

\subsubsection{Robustness}

While a calibrated reduced order model $\M{R}(\{c\})$ works well when
integrated with $c^{\ii}(t)=c(t)$, its behavior when integrated with a
different control law is unpredictable. As such, the reduced order
model can difficultly be used for estimation and optimization
purposes.\\ 
In the literature several methods are proposed for adapting reduced
order modeling for control purposes, some successful examples can be
found in \cite{Hinze2005,Rav2007,Berg2008}. However for those cases,
no calibration seems necessary for the models to work, but this is not
the case for general control problems as shown in the following.\\
The originality of the model we propose hereafter, is the combination
of multi control data sets with the calibration procedure. Such a
model is fast to build and yet remains accurate for different control
inputs. 

\subsection{Building a robust low order model}
\label{sub:cal_3}

In this section we describe a method to make the reduced order model
stable and robust.

\subsubsection{Tikhonov regularization}
\label{sub:tik}
Correctly solving (\ref{eq:adj_cal}) can be done by applying a
quasi-Newton method, initialized with $\ahb$ and $\XG_1$. It therefore
seems reasonable to solve the following regularized problem, instead
of (\ref{eq:padot}):
\begin{systa}
  \label{adot_alpha}
  \displaystyle \min_{\XC_1} \sum_{r=1}^{N_r}  \int_0^T \left
    (\dd{\ah}_r(t) - \fb_1(\ahb(t)) \cdot \XC_{r,1} \right . &-&
  {\left . \fb_2(\ahb(t)) \cdot \XG_{r,2}
    \right )}^2 \d{t} \\
  & & +  \alpha \sum_{r=1}^{N_r} \| \XC_{r,1} - \XG_{r,1} \|^2
\end{systa}
where $\alpha$ is the regularization parameter.\\
The parameter $\alpha$ can be chosen by a classical technique. We
start by plotting, for a set of values of $\alpha$ in
$[10^{-6},10^{-2}]$, the error  $\sum_r {\| \dd{a}_r - \dd{\ah}_r
  \|}^2$ versus the coefficient  variation $ {\|\XC_1 -
  \XG_1\|}^2$. This leads to a classical Tikhonov L-shaped curve of
which the corner point is optimal in  the sense that it is a good
compromise between the error on the dynamics  and the distance from
the original coefficients \cite{Hansen}. The value  of $\alpha$
corresponding to this point can be chosen to perform the  calibration
procedure.

\subsubsection{Calibrating over more than one control law}

In this paragraph we look at the changes to be made to the reduced order
model when the data set includes simulations obtained using different
control laws. Letting:
\[ \ah_r^{i,\ell} = \(\ub^{i,\ell},\phib^r \) \]
the calibration problem becomes:
\begin{eqnarray}
  \label{eq:multi_cal}
  \min_{\XU} \sum_{r=1}^{N_r}  \sum_{\ell=1}^{N_c}  \int_0^T {\(\dd{\ah}_r^{\ell}(t)-
    \fb(\ahb^{\ell}(t),c^{\ell}(t),\dd{c}^{\ell}(t)) \cdot \XU_r \)}^2 \d{t} 
\end{eqnarray}
We remark that although the size of the snapshots database is proportional to
the number of controls considered, the size of the calibration problem
remains constant. Furthermore, if $N_c > 1$ the rank deficiency
discussed for proportional feedback no longer occurs.\\
The main idea is that as the number of controls $N_c$ is increased,
although the model can become a little less precise for the reference
control, it is much more accurate for other control laws. In
the next section we show some successful examples of this method at
different Reynolds number, and for different kinds of control laws.\\
We refer to a model built using $N_c$ control laws as an $N_c$-control
model. Such a model is denoted $\M{R}_{\M{C}}$ where $\M{C} =
\{c_1,\cdots,c_{N_c} \}$.

\section{Results and discussion}

The described technique was applied in order to build a low order
model of the actuated flow around the confined square cylinder in
various configurations. We tested the prediction capabilities of the
model for two different Reynolds number, $Re = 60$ and $Re = 150$,
with precomputed and feedback control laws. In particular we built
different models with one and more control laws and we analyzed their
predictions with different controls.\\ 
In all the examples presented in the following, actuation is started
only once the flow is fully developed. With the control turned on the
simulation is performed for about seven vortex shedding cycles, and
$N_t \approx 200$ snapshots are saved. $T \simeq 50$ is the
non-dimensional duration of the time interval. The number of POD modes
retained for the reduced order model is $N_r=40$ for the case $Re=60$
and $N_r=60$ for the case $Re=150$.\\ 
We measure the accuracy of the model $\M{R}(\M{C})$ in the following
way: 
\begin{itemize}
\item Time coefficients dynamics:\\
For a given value of $r$, plot  $a_r(t)$, solution of $\M{R}({\M{C}})$ with input $c^{\ii}(t)$, against
$\ah_r(t)$, projection of the full order solution onto the POD basis
$\phib({\M{C}})$. In the examples $r=3$ is usually chosen because it was the
mode for which the differences between models were the most remarkable.
\item Computation of the integration error:
  \[ \M{E}({\M{C}},c^{\ii}) = \|\ab-\ahb\| / \|\ahb\| \]
where $\displaystyle \|\ab\| =  \int_0^T \sum_r a_r^2 (t) \d{t}$
\end{itemize}
In the examples with feedback laws we use only one sensor placed in
the cylinder wake. Choosing the center of the cylinder as the origin
of a coordinate system, we denote $\xb_s = (x_s, y_s)$ the position of
the sensor. The integration error $\M{E}^{\ff}({\M{C}},\Kb^{\ii})$ is
measured in the same way as for the non-feedback case. \\
Our first goal is that the model should be able to reproduce the DNS
data to which is was fitted, we therefore expect
$\M{E}({\M{C}},c^{\ii})$ to be small if $c^{\ii} \in \M{C}$. Our
second goal is that the model be robust to parameter variation. As the
difference between $c^{\ii}(t)$ and the controls in $\M{C}$ increases,
the error $\M{E}({\M{C}},c^{\ii})$ grows. We seek a model for which
this growth rate is as low as possible. 

\subsection{Divergence of a classical Reduced Order Model}
\label{res:pbs}

\begin{figure}[!t]
  \centering 
  \hspace{-3.45cm}
  \includegraphics[width=3.5cm,height=3.5cm]{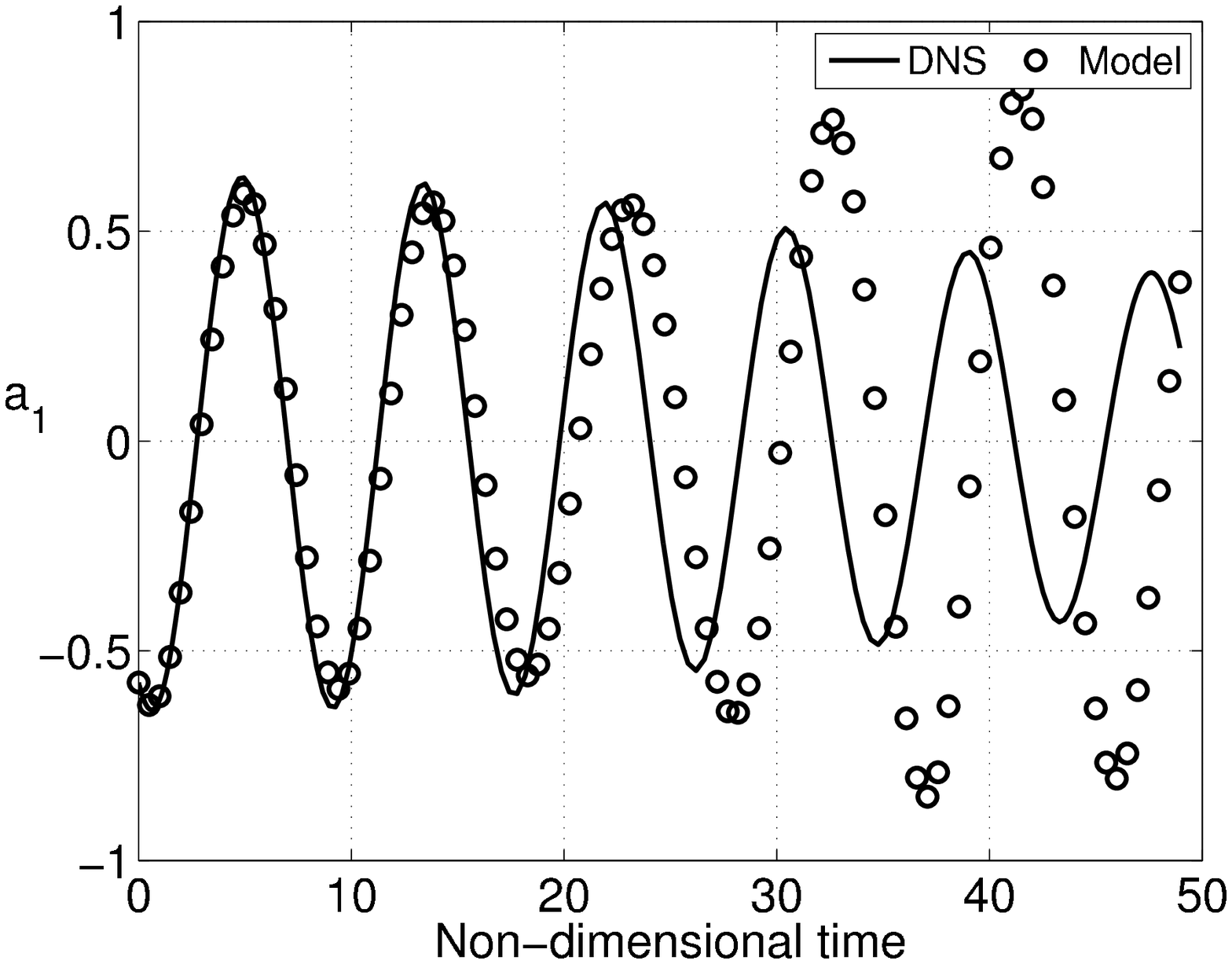}\\
  \vspace{-5.4cm}
  \hspace{4.cm}
  \includegraphics[width=4.cm,height=7.25cm]{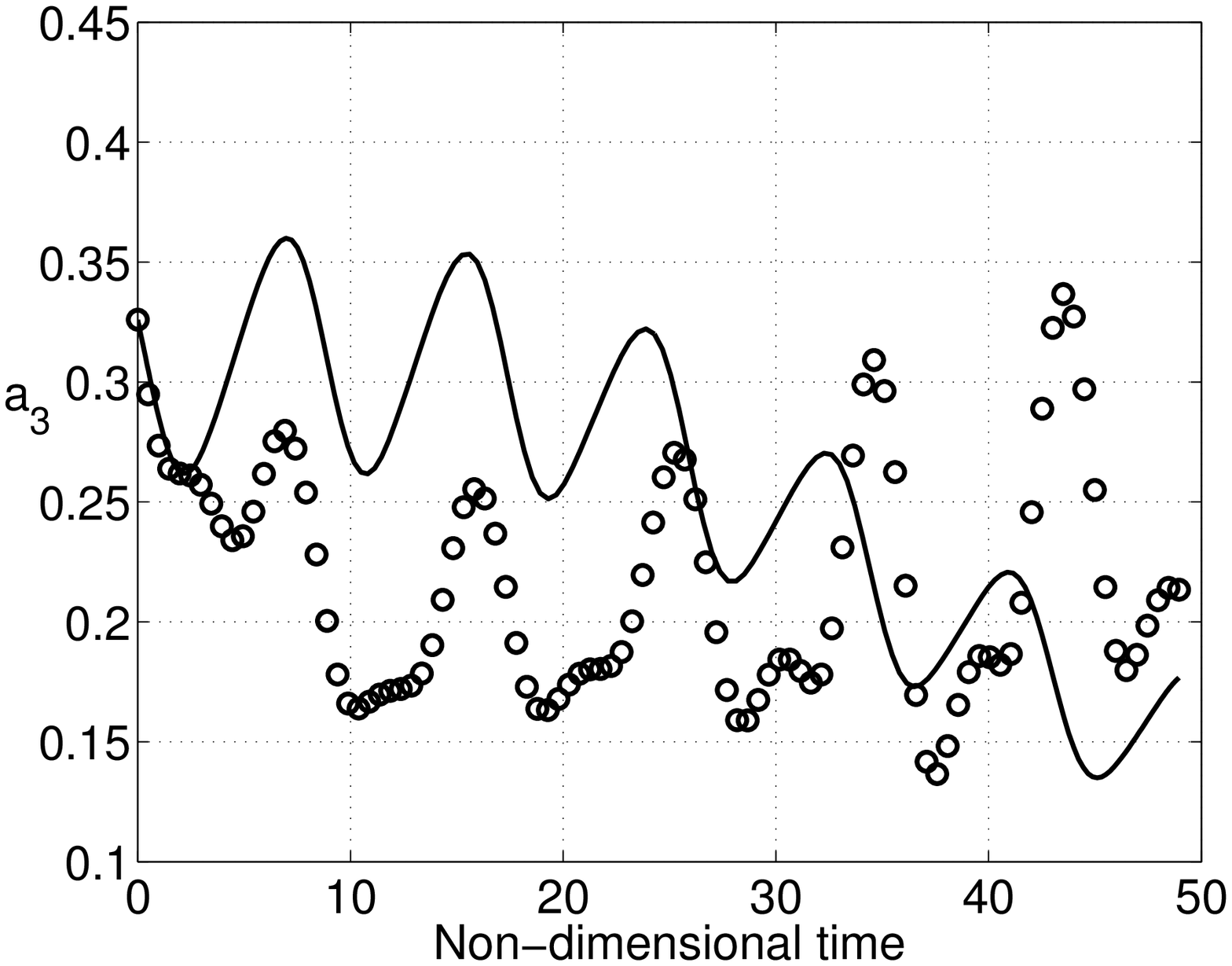}\\
  \vspace{-2.cm}
  \caption{Projection of the DNS simulations onto POD
  modes vs. integration of the dynamical system (\ref{eq:romf}) with $\XU=\XG$ }
  \label{fig:exple_fbck}
\end{figure}

\begin{figure}[!t]
  \centering 
  \includegraphics[width=3.5cm,height=3.5cm]{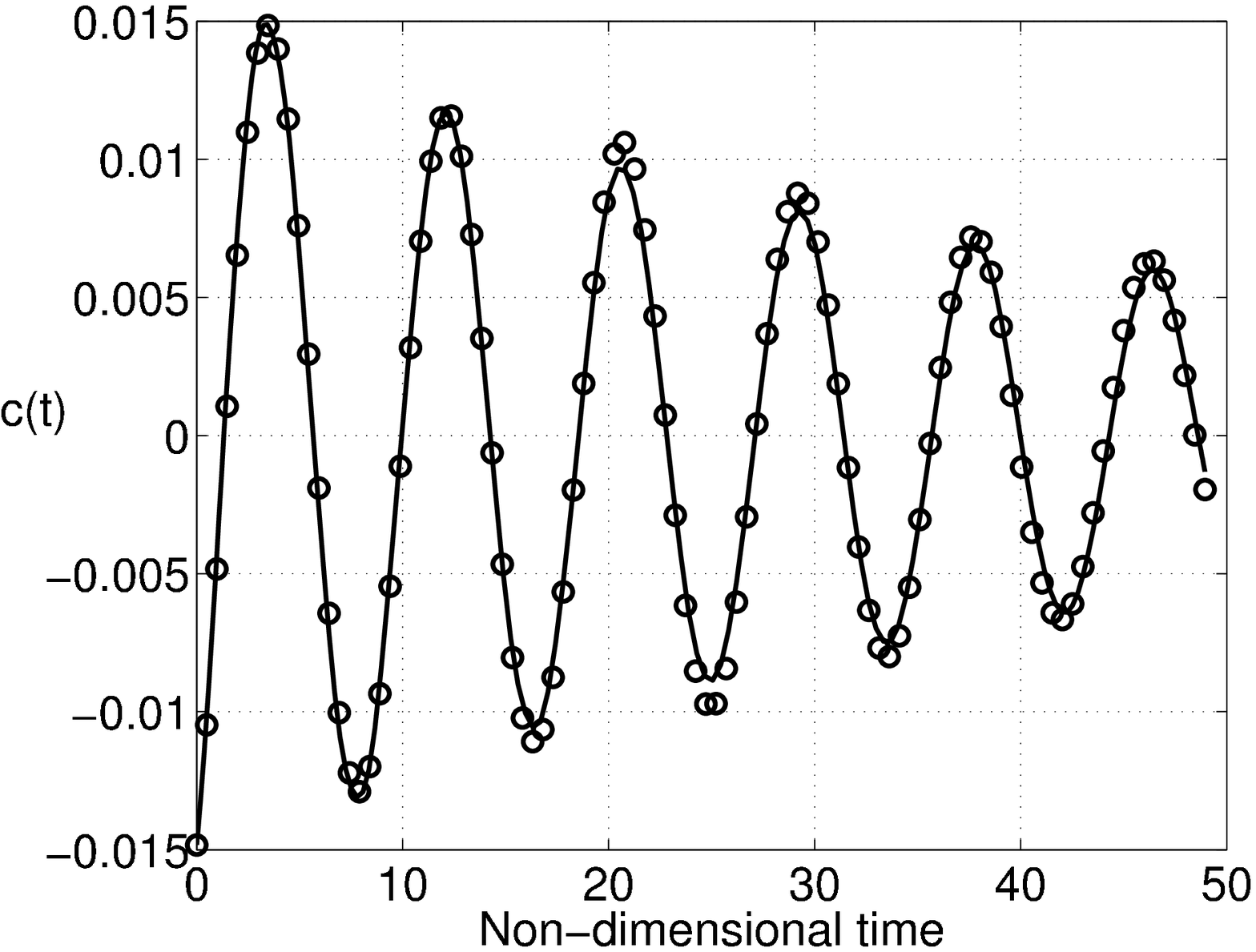}
  \includegraphics[width=3.5cm,height=3.5cm]{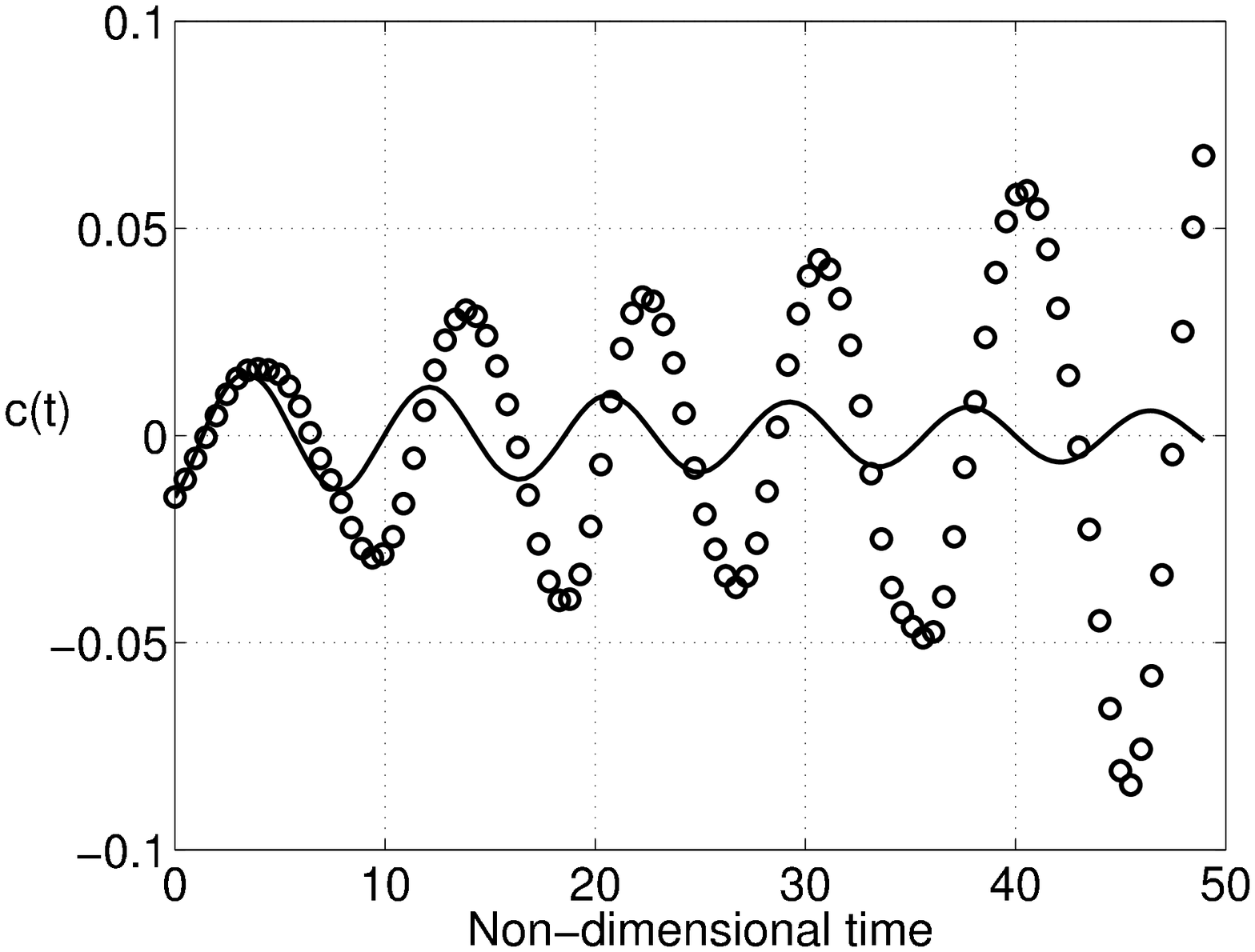}\\
  \caption{$c_1$ (continuous line) versus $c^{\ff}$, when the model is
    calibrated (left) and when it is not (right)}
  \label{fig:ctrl_fbck}
\end{figure}

\begin{figure}[!t] 
  \centering 
  \includegraphics[width=4.25cm,height=6.5cm]{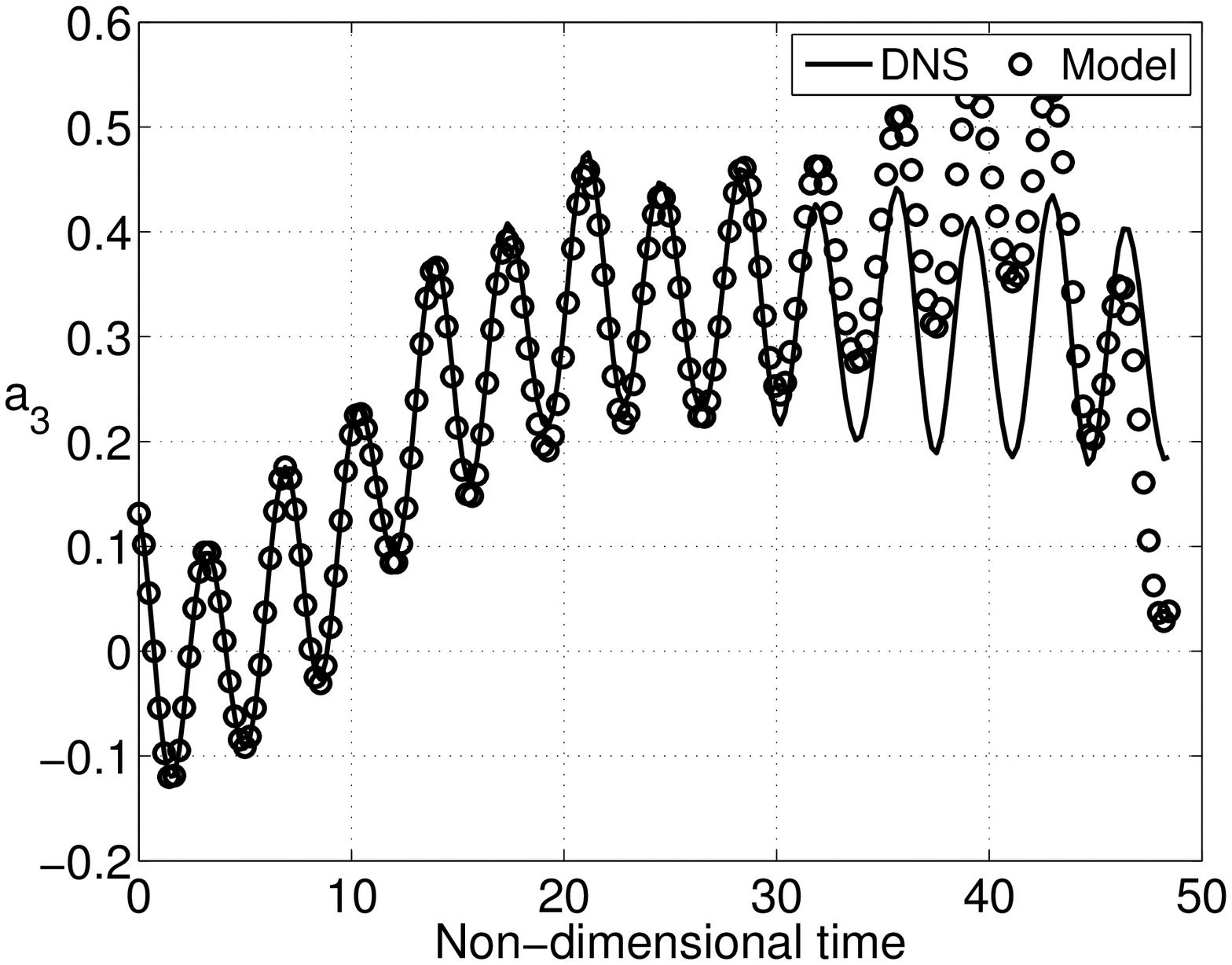}
  \includegraphics[width=4.25cm,height=6.5cm]{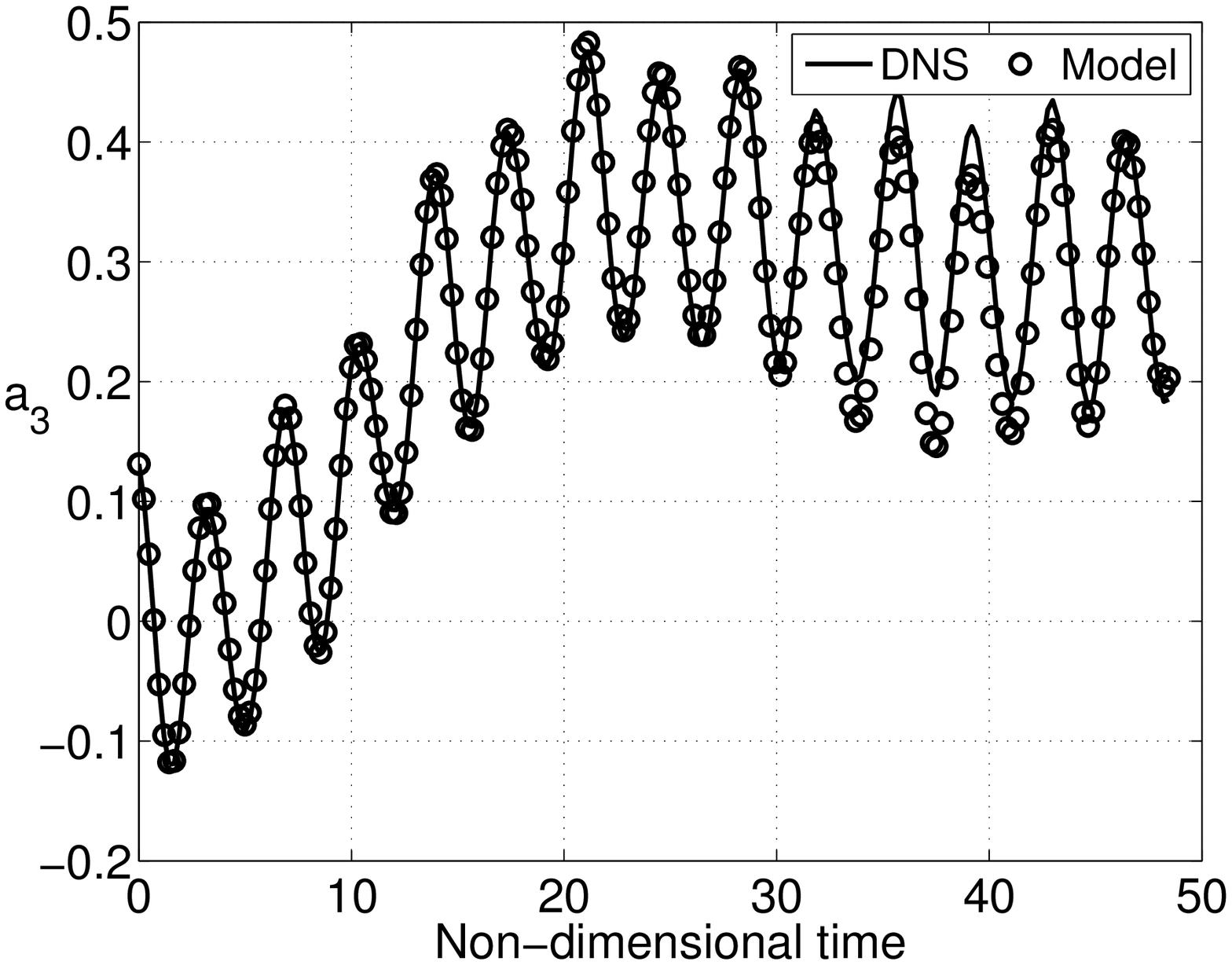}\\
  \vspace{-3.cm}
  \includegraphics[width=4.25cm,height=6.5cm]{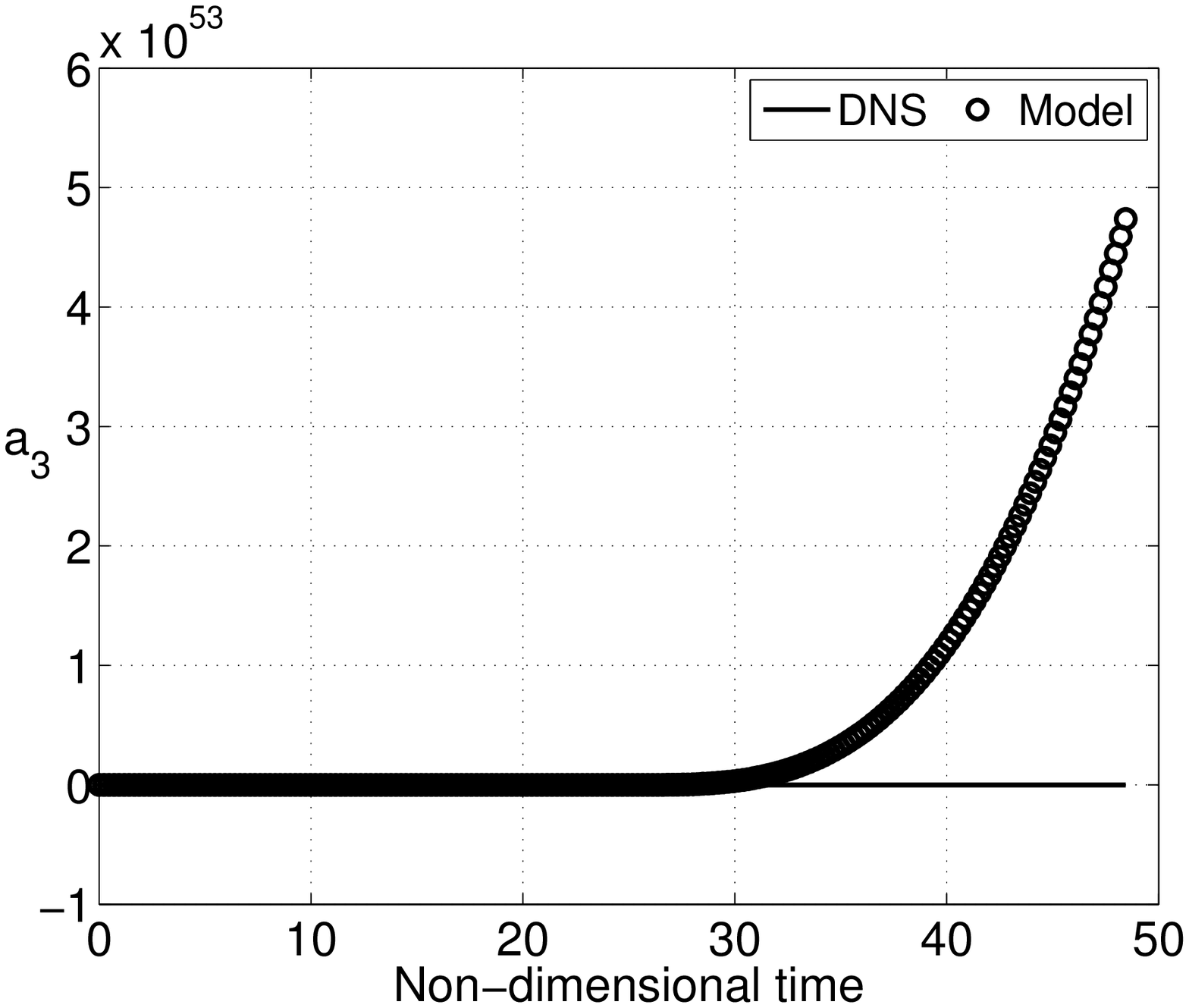}
  \includegraphics[width=4.25cm,height=6.5cm]{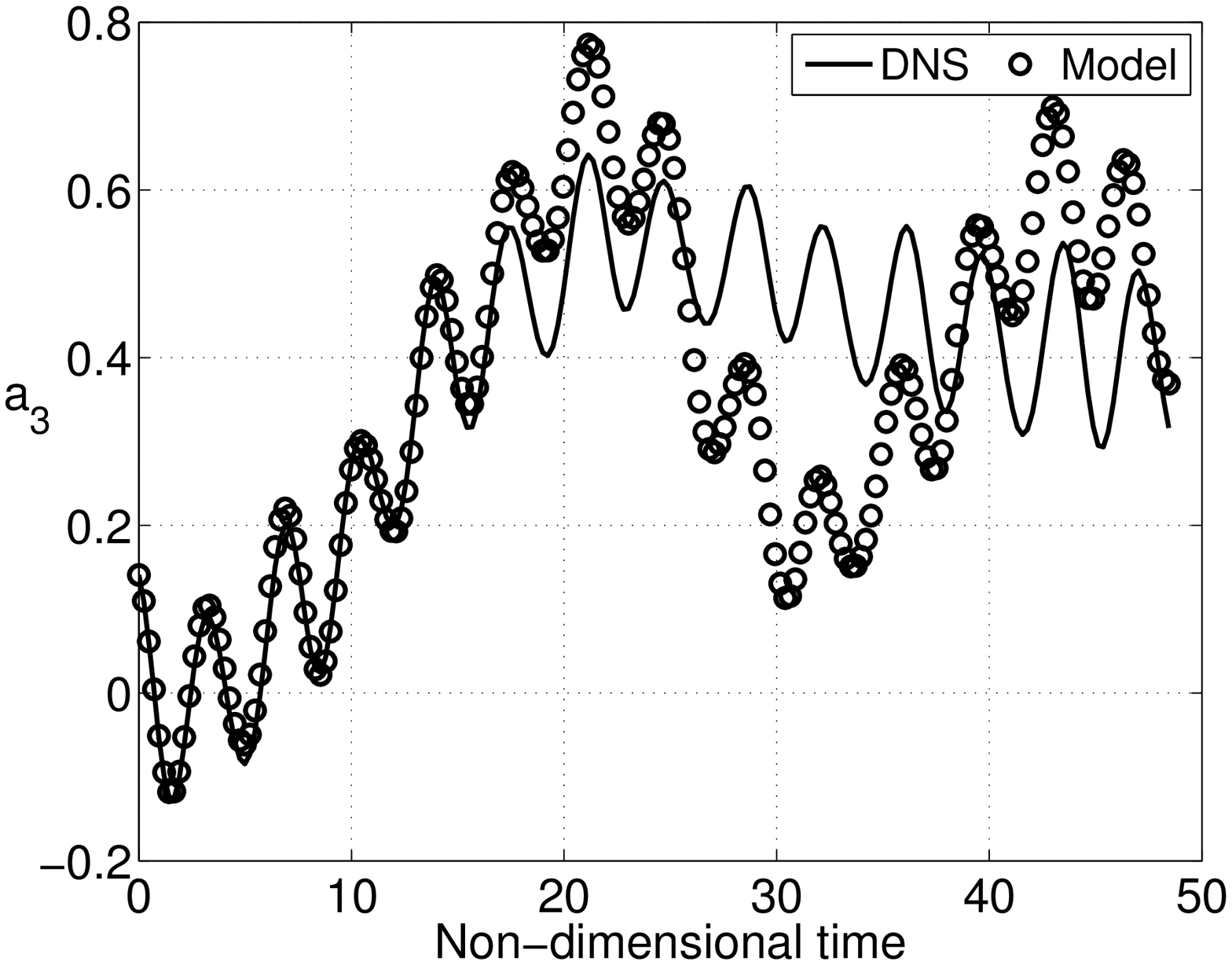}\\
  \vspace{-1.5cm}
  \caption{$a_3$ DNS (continuous line) with $\Kb=0.8$ (top) and $\Kb=1.3$
    (bottom) versus $a_3$ obtained when the model is
    calibrated with $\alpha = 1.6 * 10^{-6}$ (left) and when $\alpha
    = 10^{-3}$ (right)}
  \label{fig:integr_fbck_alpha}
\end{figure}

A simulation at $Re=60$ was performed using feedback control with a
sensor placed at $(x_s,y_s)=(0.7,0.0)$ and $\Kb=1$. We denote $c(t)$
the control law obtained at the end of simulation. \\
We compare the results obtained with the POD Galerkin model
(\ref{eq:gal_rom}) and with the calibrated model $\M{R}(\{c\})$ 
(see system (\ref{eq:rom}) for model formulation). 
The model integration error $\M{E}({\{c\}},c)$ is equal to $23\%$ for
the non-calibrated model, and to $0.136\%$ for the calibrated model.\\
For a feedback model, the difference is even more important. We
integrated the feedback system (\ref{eq:romf}) with $\Kb=1$, once with
$\XU$ obtained by Galerkin projection, and once with $\XU$ calibrated
as described in \ref{sub:cal_fbck}. We obtained an integration error
$\M{E}^{\ff}(\{\ch\},1)$ of $117\%$ in the first case, against an
error of $4\%$ in the other. An example of the errors in terms of time
dynamics that the non-calibrated model can produce are shown in
Fig.\ref{fig:exple_fbck}. 
In Fig.\ref{fig:ctrl_fbck} we plot the control law $c^{\ff}(t)$ computed when
integrating the feedback model, and on the same figure, the original
control law $c(t)$. Results for the non-calibrated case are plotted on
the right: the distance between $c^{\ff}(t)$ and $c(t)$ increases with
time, meaning that at each time step, new errors are added to
the model. Calibration is therefore all the more essential when considering
feedback control. \\
In order to calibrate, regularization is needed to get well-conditioned
inverse problems as shown in the following. However, the
choice of the  parameter $\alpha$ is not an easy one. \\
For example, we performed a simulation at $Re=150$ using a feedback
control with a sensor placed at $(x_s,y_s)=(0.7,0.0)$ and
$\Kb=0.8$. The calibration described in \ref{sub:tik} was performed
with $\alpha \approx 0$. This led to an ill-conditioned system to
solve and to a model that was not very accurate, and not robust at all
to parameter variations. The effect of $\alpha$ on model results is
shown in Fig.\ref{fig:integr_fbck_alpha}. The two top figures show the
third modal coefficient obtained by projection and by integrating the
model with $\Kb=0.8$. On the left, we plot the results obtained when
the model was built with $\alpha = 1.6*10^{-6}$: at the end of the
time period the model diverges from the DNS results. With a higher
value,  $\alpha = 10^{-3}$, this problem no longer occurs, as shown on
the right. The same test was then performed with a different value of
$\Kb$ in order to see the models capacity to predict dynamics to which
it was not fitted. The results are shown in the same figure:
divergence was immediate for a low value of $\alpha$, whereas for a
higher value, the model, although not accurate, was at least stable.  
It appears that, when using an higher regularization parameter, the
calibration system is well conditioned, the model more accurate and more
stable when integrated with a different control law to those used for
calibration. In the following the parameter $\alpha$ is determined
using the L-method with the restriction that any values of $\alpha$
below a certain threshold are excluded. 

\subsection{Testing model robustness}

In this section we present the improvements  brought to model
robustness by introducing calibration over several control laws. For
both Reynolds numbers, $Re=60$ and $Re=150$, the same experiment was
performed:

\subsubsection*{Step 1 : Build 1- , 2- and 3-control models}
%\textbf{Step 1 : Build 1- , 2- and 3-control models :}\\
We started by choosing three control laws which we denote $c_1(t)$, $c_2(t)$ and
$c_3(t)$. For each control we performed a simulation of the
Navier-Stokes equations, saving $200$ snapshots for each
simulation. We then defined seven control sets: 
\[\mbox{Three 1-control sets: } \M{C}^1 = \{c_1\}, \M{C}^2 =
\{c_2\},\M{C}^3 = \{c_3\} \]
\[\mbox{Three 2-control sets: } \M{C}^4 = \{c_1,c_2\}, \M{C}^5 =
\{c_1,c_3\}, \M{C}^6 = \{c_2,c_3\}\]
\[\mbox{One 1-control set: } \M{C}^7 = \{c_1,c_2,c_3\} \]
For each control set $\M{C}^i$, we computed a POD basis
$\phib({\M{C}^i})$ as described in section \ref{sub:podbasis} and a
calibrated reduced order model $\M{R}({\M{C}}^i)$ by solving problem
(\ref{eq:multi_cal}).\\
In the following we refer to  $c_1(t)$, $c_2(t)$ and
$c_3(t)$ as the \textsl{model control laws}.

\subsubsection*{Step 2 : Run the model with different control laws}
%\textbf{Step 2 : Run the model with different control laws}\\
We next chose several other control laws which we denote
$c_j^{test}(t)$. Each of these \textsl{test control laws} was used as
input for the Navier-Stokes equations, and for the seven reduced order
models $\M{R}({\M{C}^i})$ described above. The snapshots from the
Navier-Stokes simulations were projected onto the seven POD bases
$\phib({\M{C}^i})$. This procedure made it possible to compute the
model integration errors $\M{E}_i^j = \M{E}({\M{C}^i},c_j^{test})$,
and compare the efficiency of each model.\\ 
For measuring model robustness, it is useful to have some idea of how
much the dynamics we are trying to predict, differ from those included
in the model. We therefore need to find a way, for each model, to
measure the distance between the $N_t \times N_c$ snapshots that were
used to build it, and the $N_t$ snapshots obtained using a
\textsl{test} control law. To do this we proceed in the following way
: if the control set $\M{C}^i$ is composed of $N_c$ control laws, then
the  \textsl{distance} between the simulations associated to
$\M{C}^i$, and the one obtained using $c_j^{test}(t)$, is defined as:
\[ \Delta_i^j = \frac{1}{N_c} \sum_{l=1}^{N_c} \left ( \|\ahb^l -
    \ahb^j \| / \| \ahb^l \| \right )  \]
where the terms $\ahb^n$ ($n=j$ or $n=1\cdots l$) result from
projecting the snapshots onto the POD basis  $\phib({\M{C}^i})$.\\ 
The results are plotted for in Fig. \ref{fig:multi60} and
Fig. \ref{fig:multi150}. For each value of model  $i$, the model
integration error $\M{E}_i^j$ is plotted versus the  \textsl{distance}
$\Delta_i^j$.  We note that the three controls used to build the
models were in fact included in the test set, which explains why there
are 3 points at $\Delta_i^j = 0 $.\\  

\subsection{Results for $Re = 60$}
\label{res:re60}

\begin{figure}[!t]
  \centering 
  \vspace{-1.5cm}
  \hspace{0.25cm}
  \includegraphics[width=3.9cm,height=6.5cm]{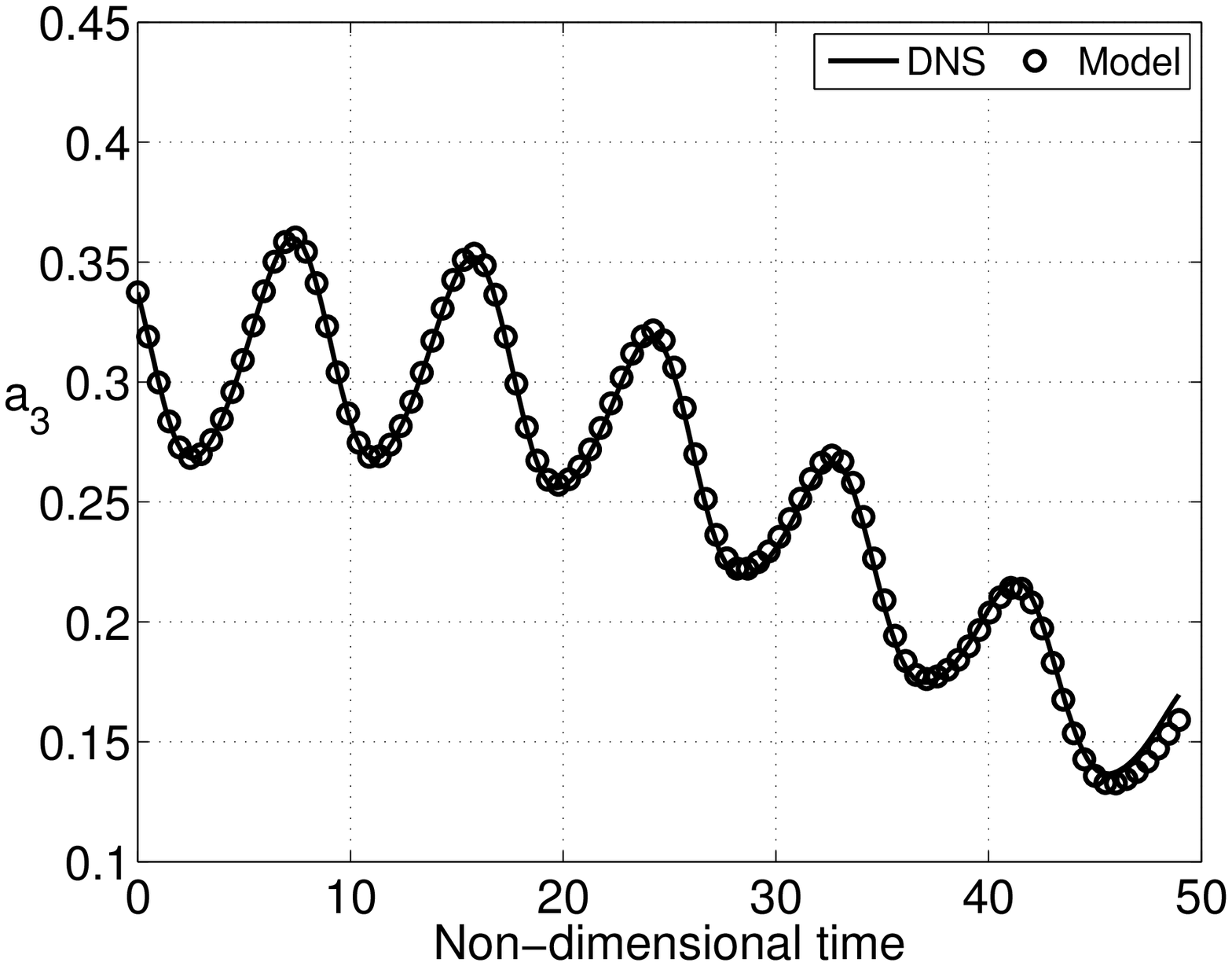}
  \hspace{-0.5cm}
  \includegraphics[width=3.9cm,height=6.5cm]{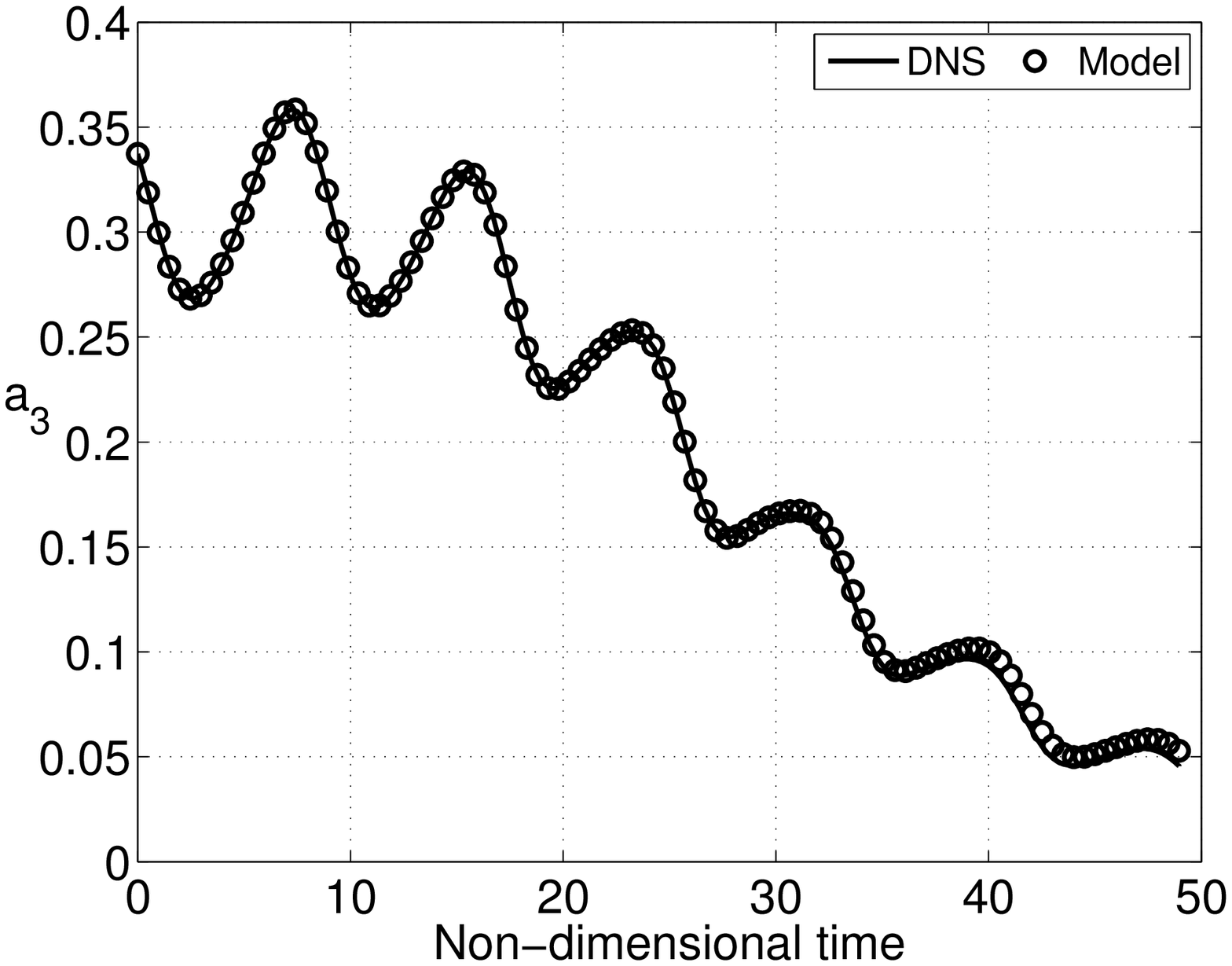}
  \hspace{-0.5cm}
  \includegraphics[width=3.9cm,height=6.5cm]{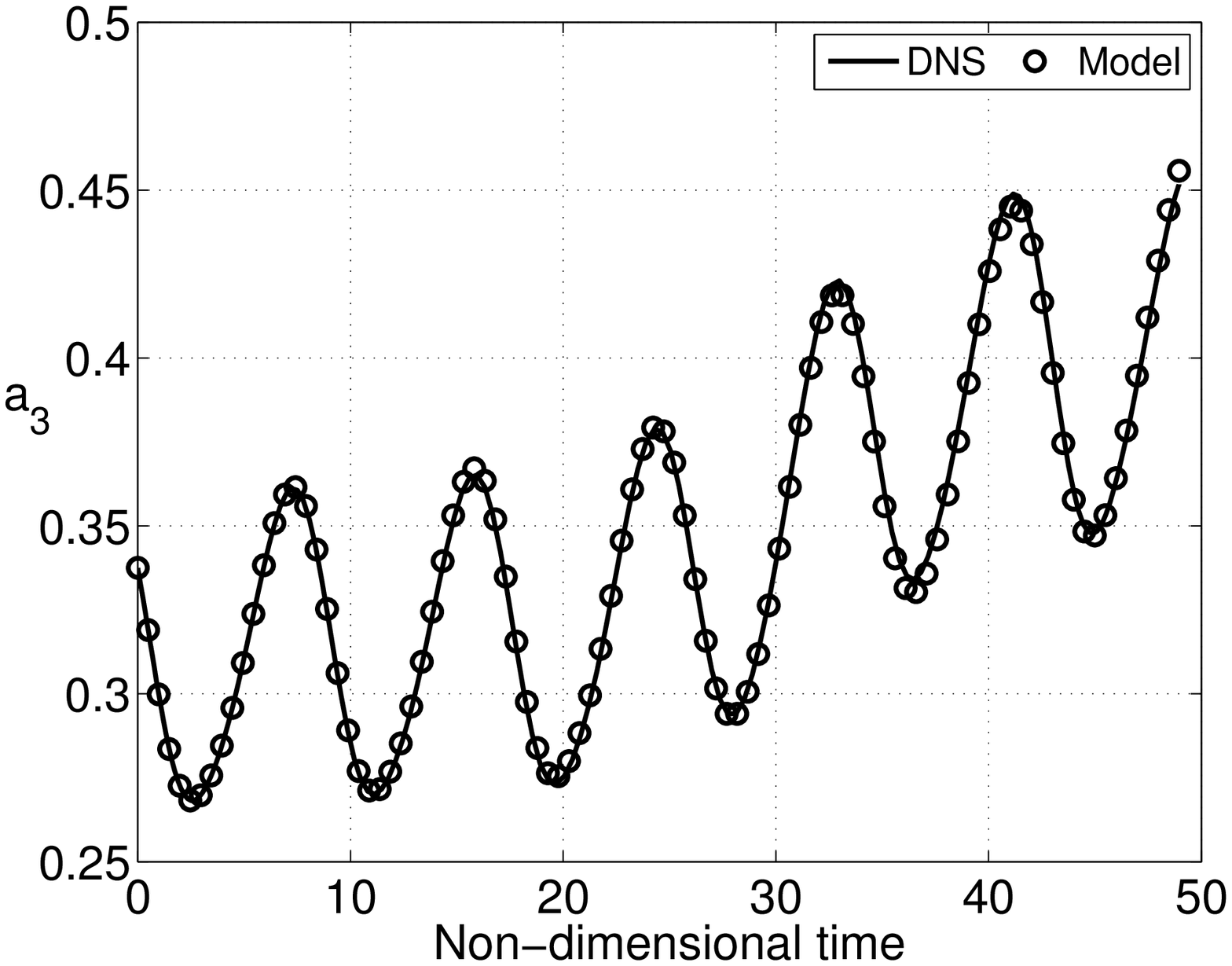} \\ 
  \vspace{-1.5cm}
    \caption{Control laws used to build the models (top); $a_3$ DNS
    (continuous line) vs prediction by 3-control model (symbols)}
  \label{fig:cal60}
\end{figure}

\begin{figure}[h]
  \centering 
  \includegraphics[width=3.9cm,height=6.5cm]{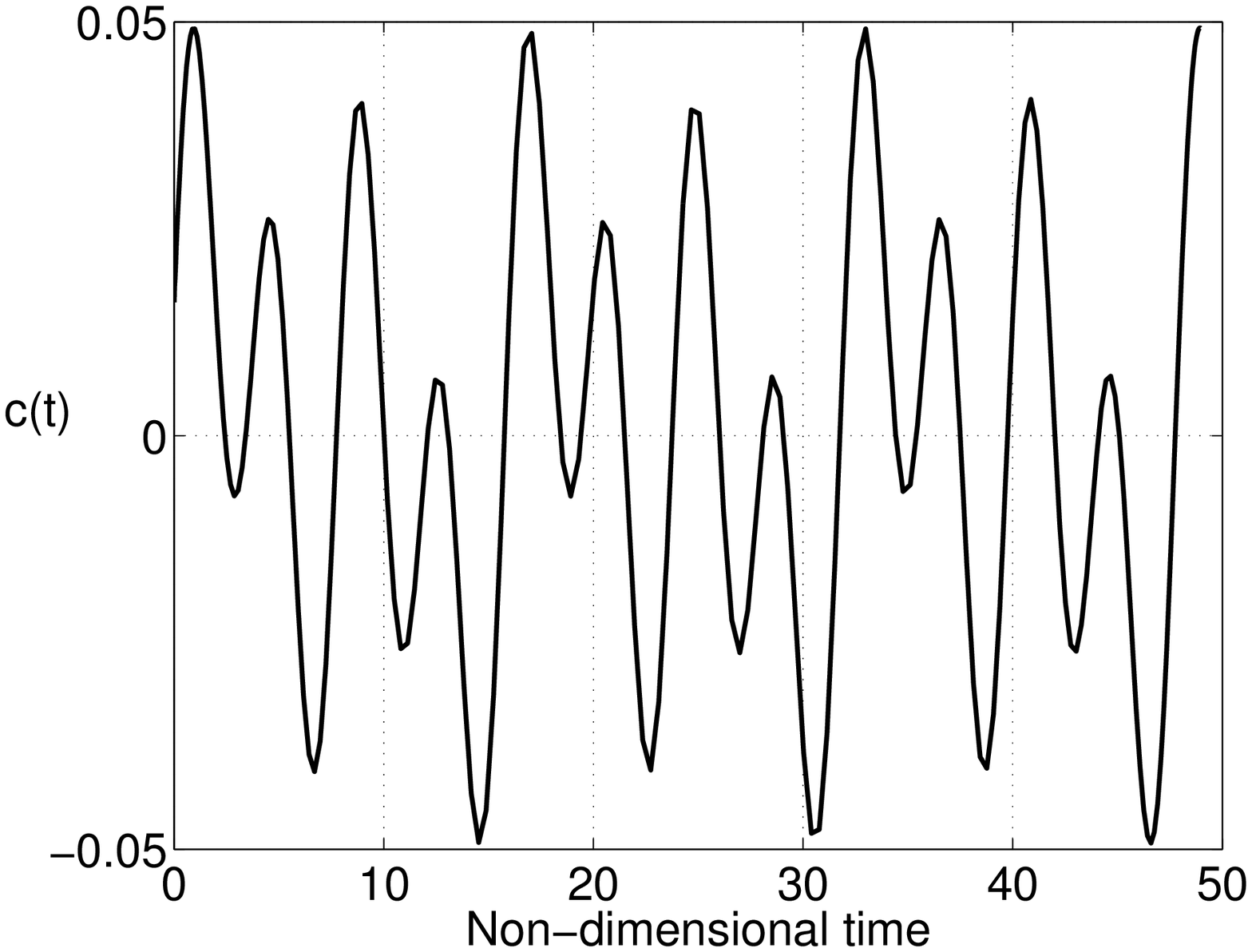}
  \hspace{-0.5cm}
  \includegraphics[width=3.9cm,height=6.5cm]{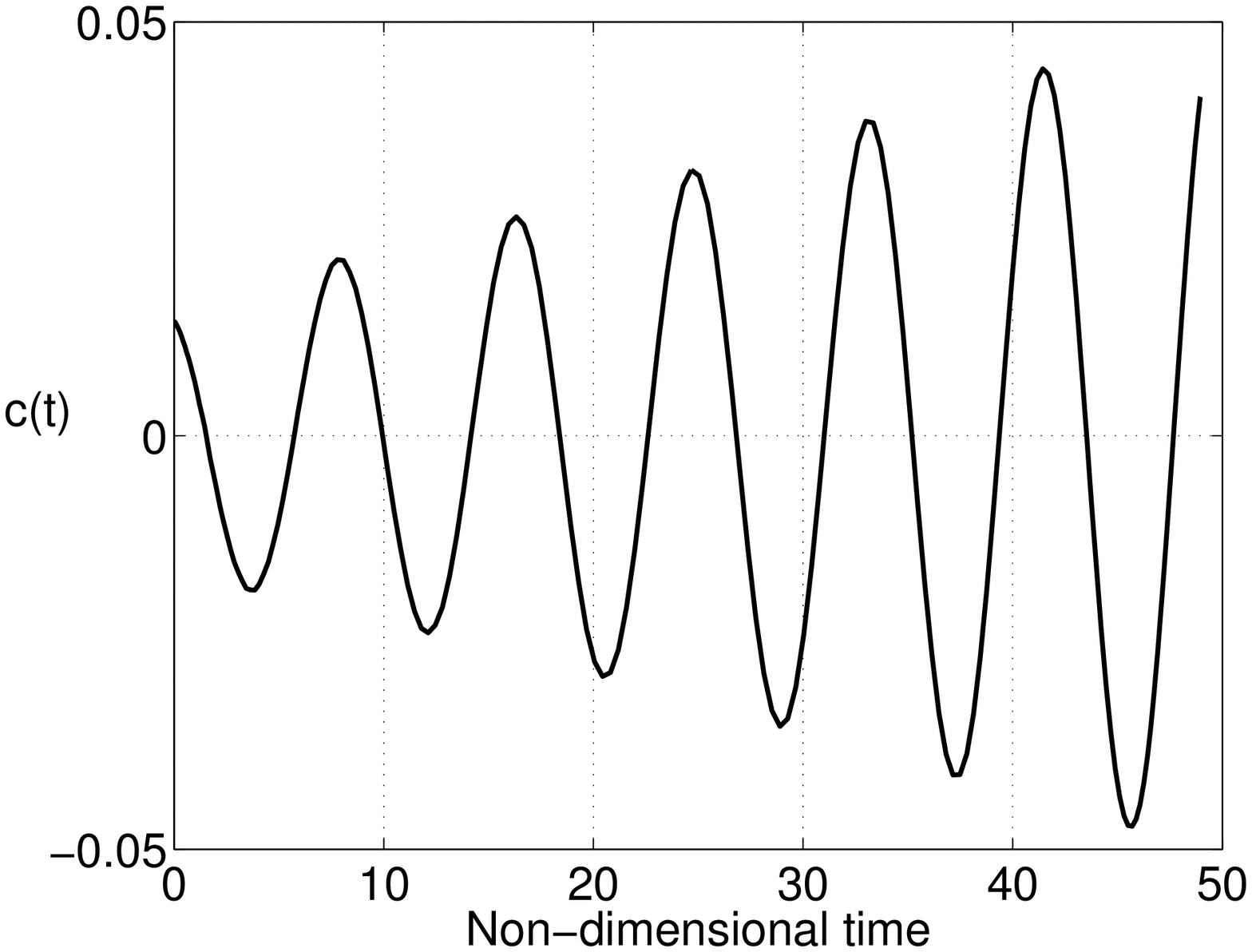}
  \hspace{-0.5cm}
  \includegraphics[width=3.9cm,height=6.5cm]{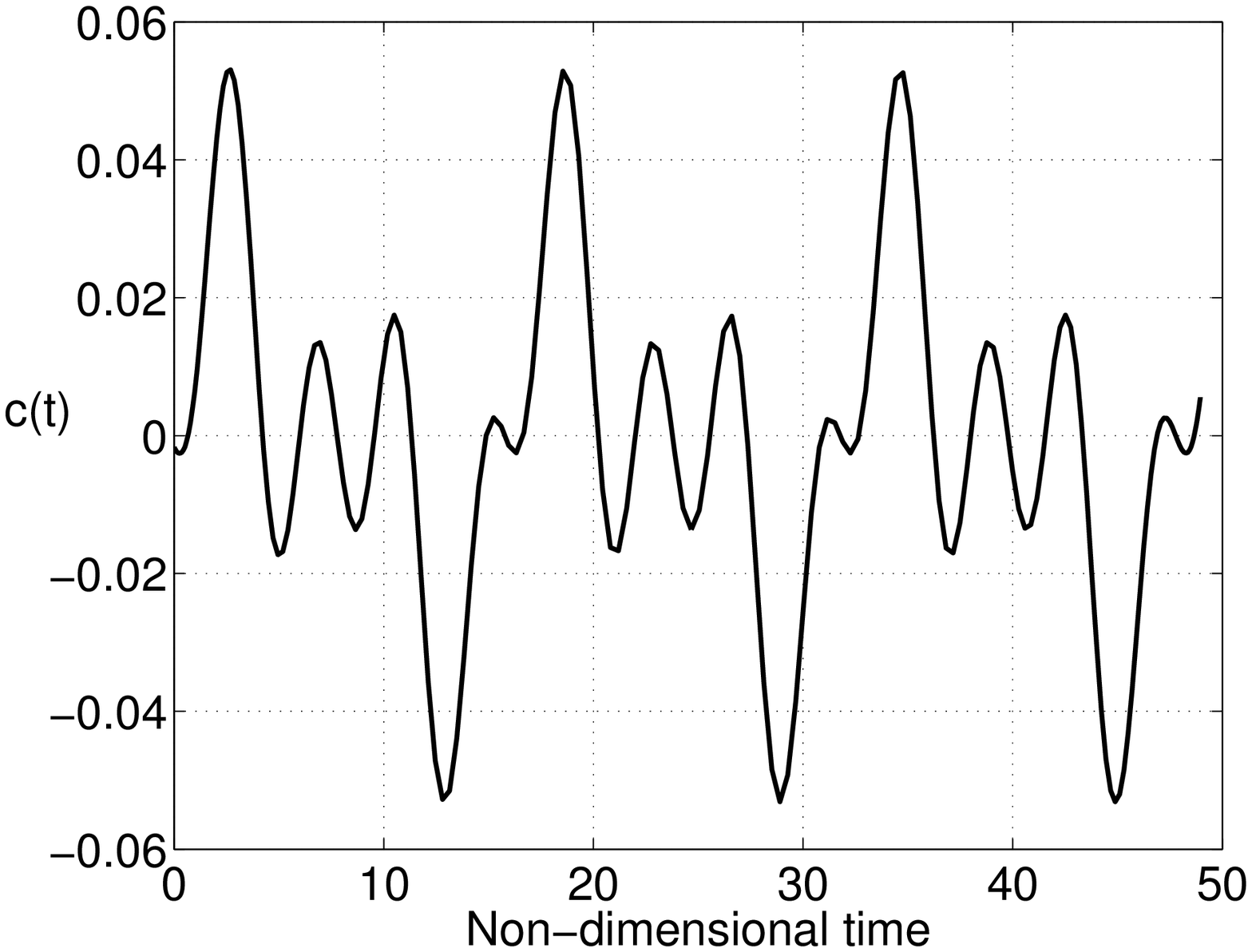}\\
  \vspace{-2.8cm}
  \hspace{0.1cm}
  \includegraphics[width=3.95cm,height=6.5cm]{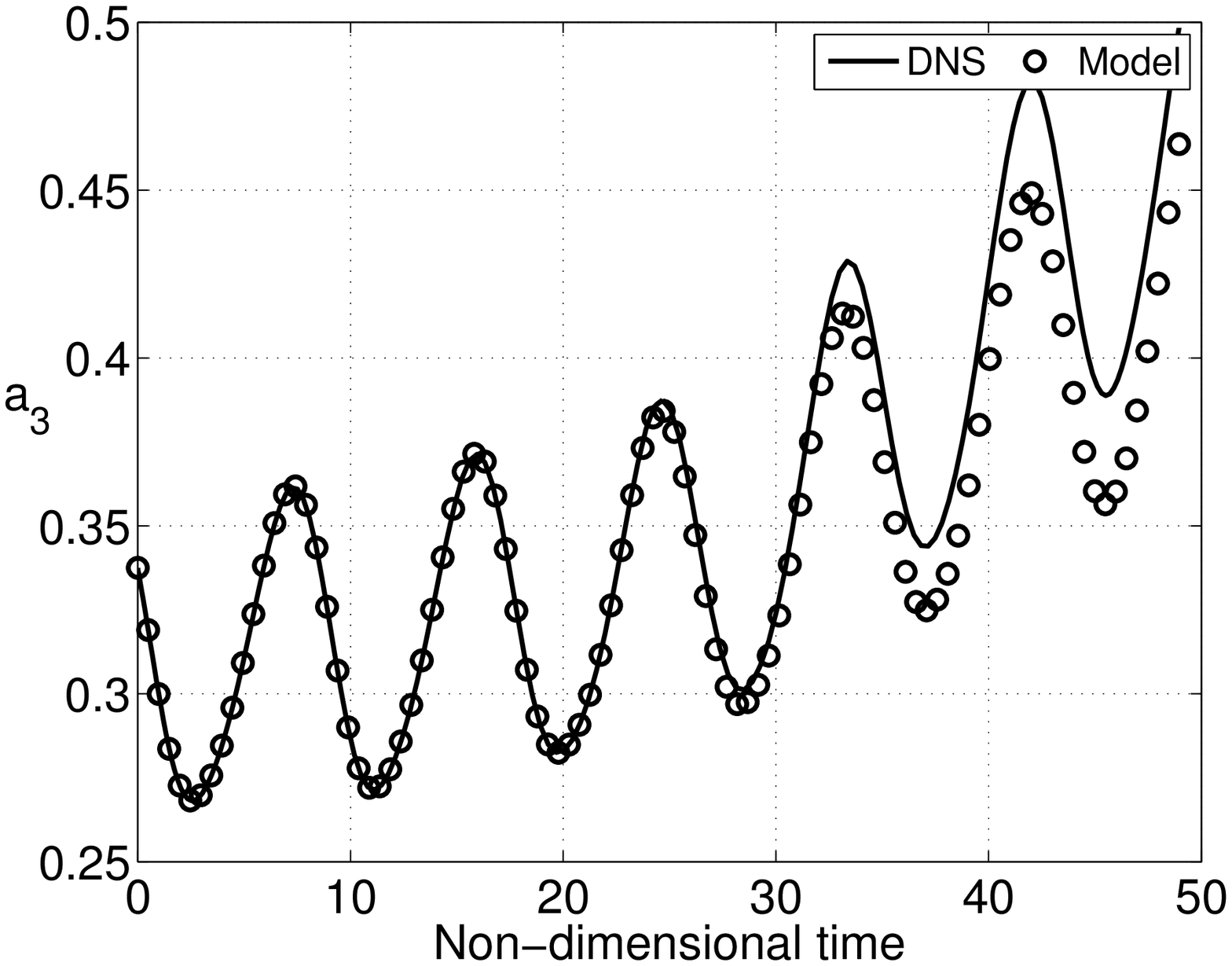}
  \hspace{-0.5cm}
  \includegraphics[width=3.95cm,height=6.5cm]{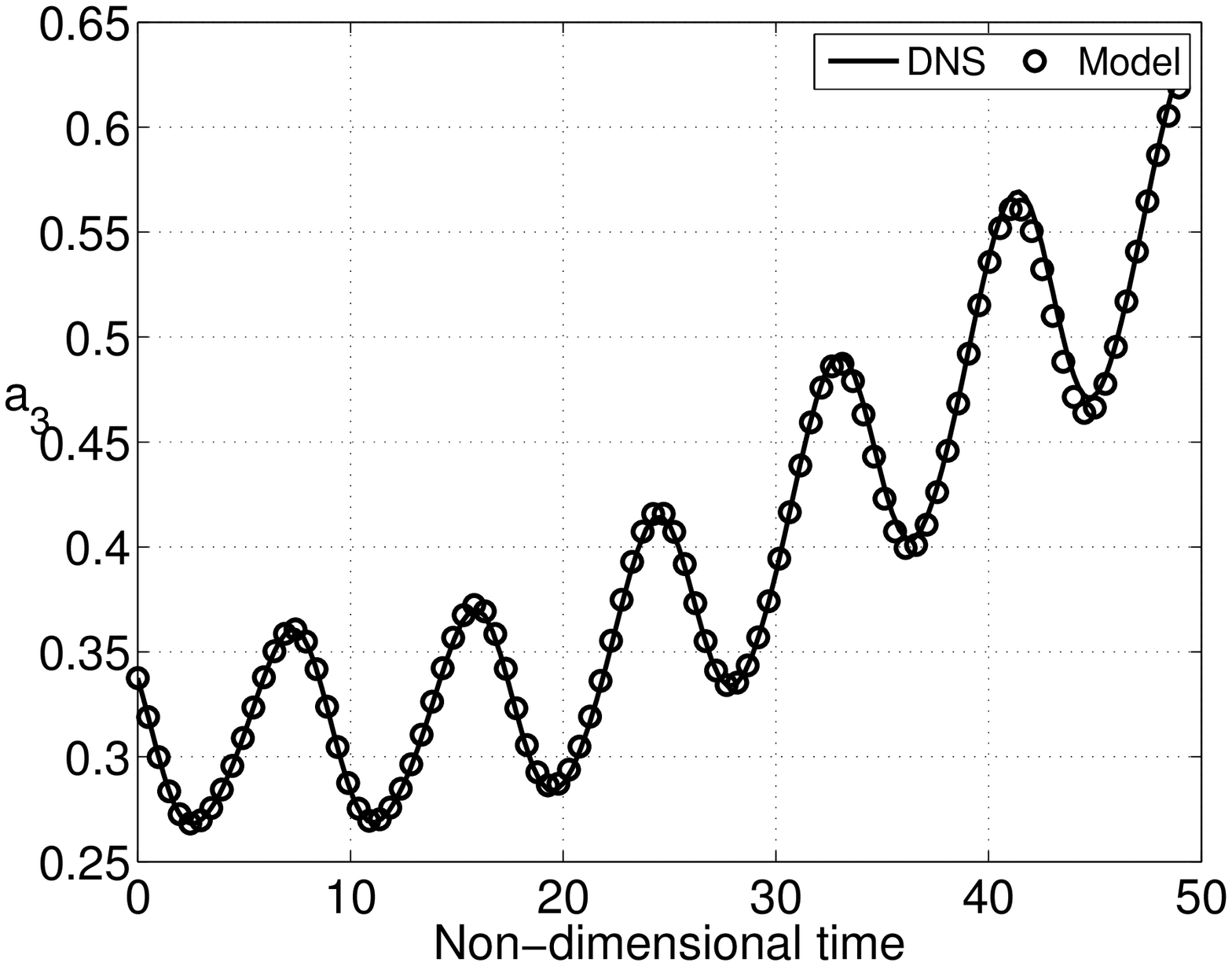}
  \hspace{-0.5cm}
  \includegraphics[width=3.95cm,height=6.5cm]{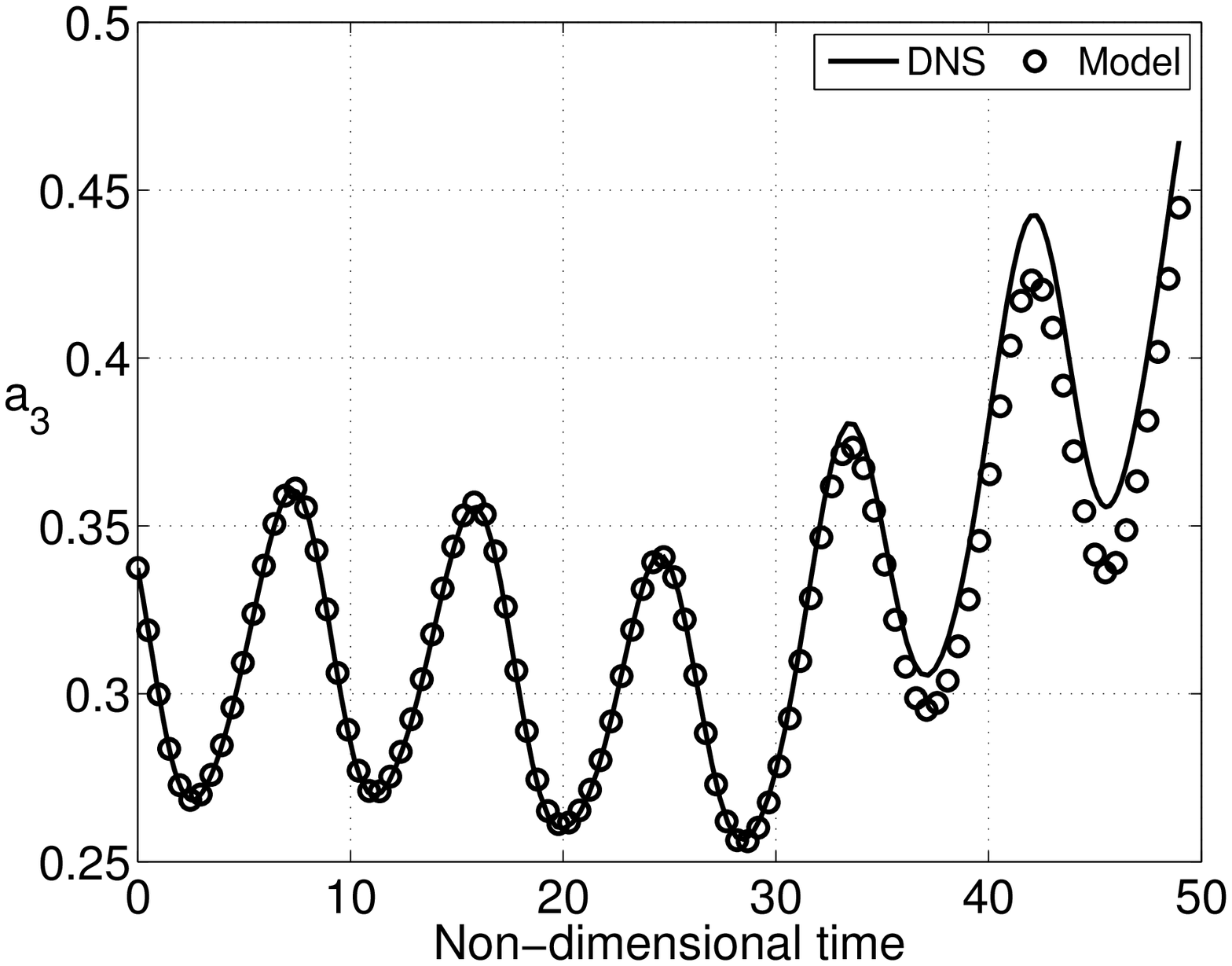}
  \vspace{-1.5cm}
  \caption{Control laws and time coefficients used for testing the model}
  \label{fig:test60}
\end{figure}

\begin{figure}[]
  \centering 
  \vspace{-2.5cm}
  \includegraphics[width=8.5cm,height=12.5cm]{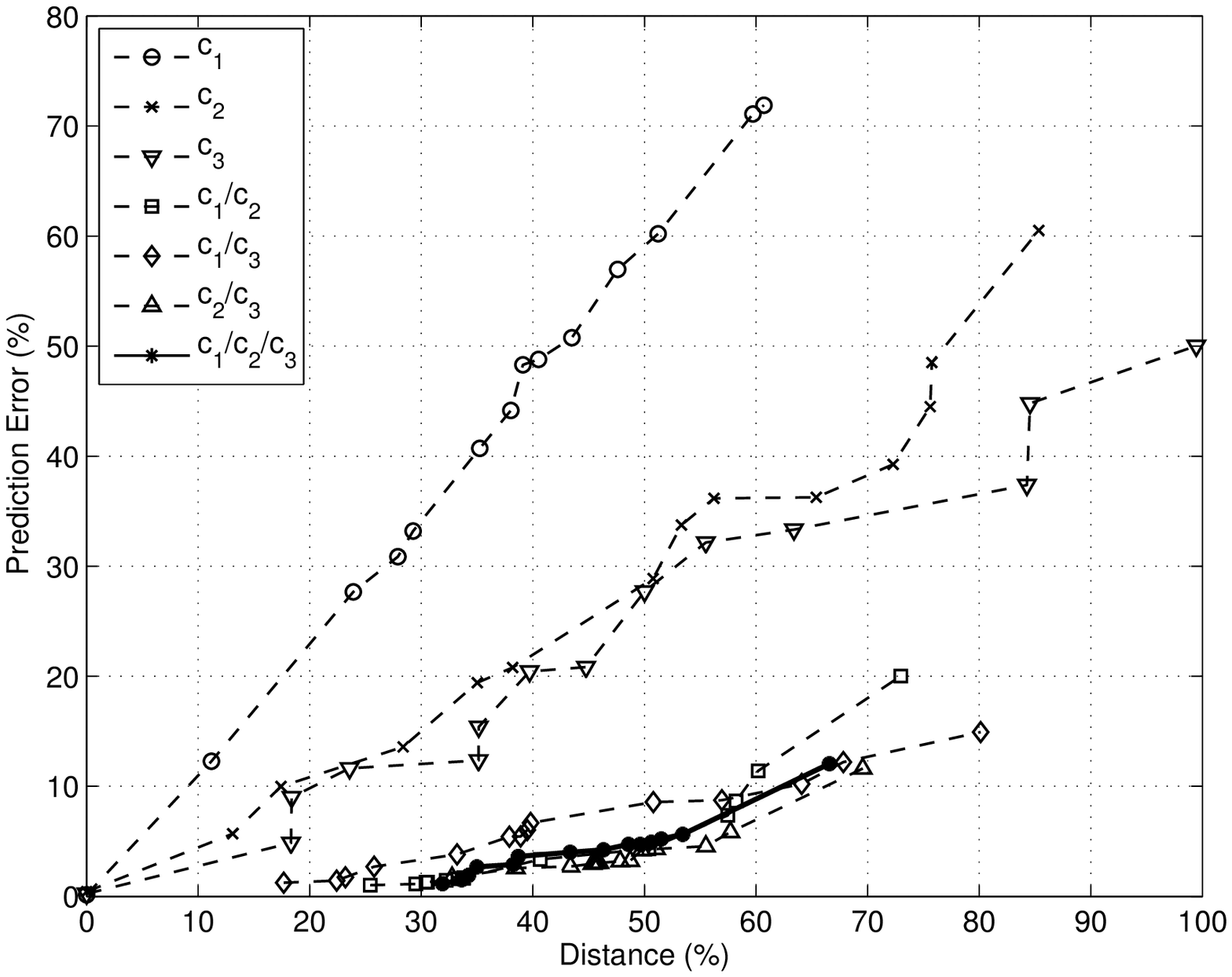}
  \vspace{-3.cm}
  \caption{Prediction errors obtained using 1-control, 2-control and
    3-control models}
  \label{fig:multi60}
\end{figure}

In Fig.\ref{fig:cal60} we plot the control laws used to build the
models. For each control law we plot the third modal
coefficient $\ah_3(t)$ to give an idea of the dynamics induced. The
figure also shows the prediction for this coefficient given by the
3-control model $\M{R}({\M{C}^7})$. The model results are accurate:
the reduced order model was successfully calibrated to fit several
dynamics. 
\begin{figure}[!t]
  \centering 
  \includegraphics[width=3.5cm,height=3.5cm]{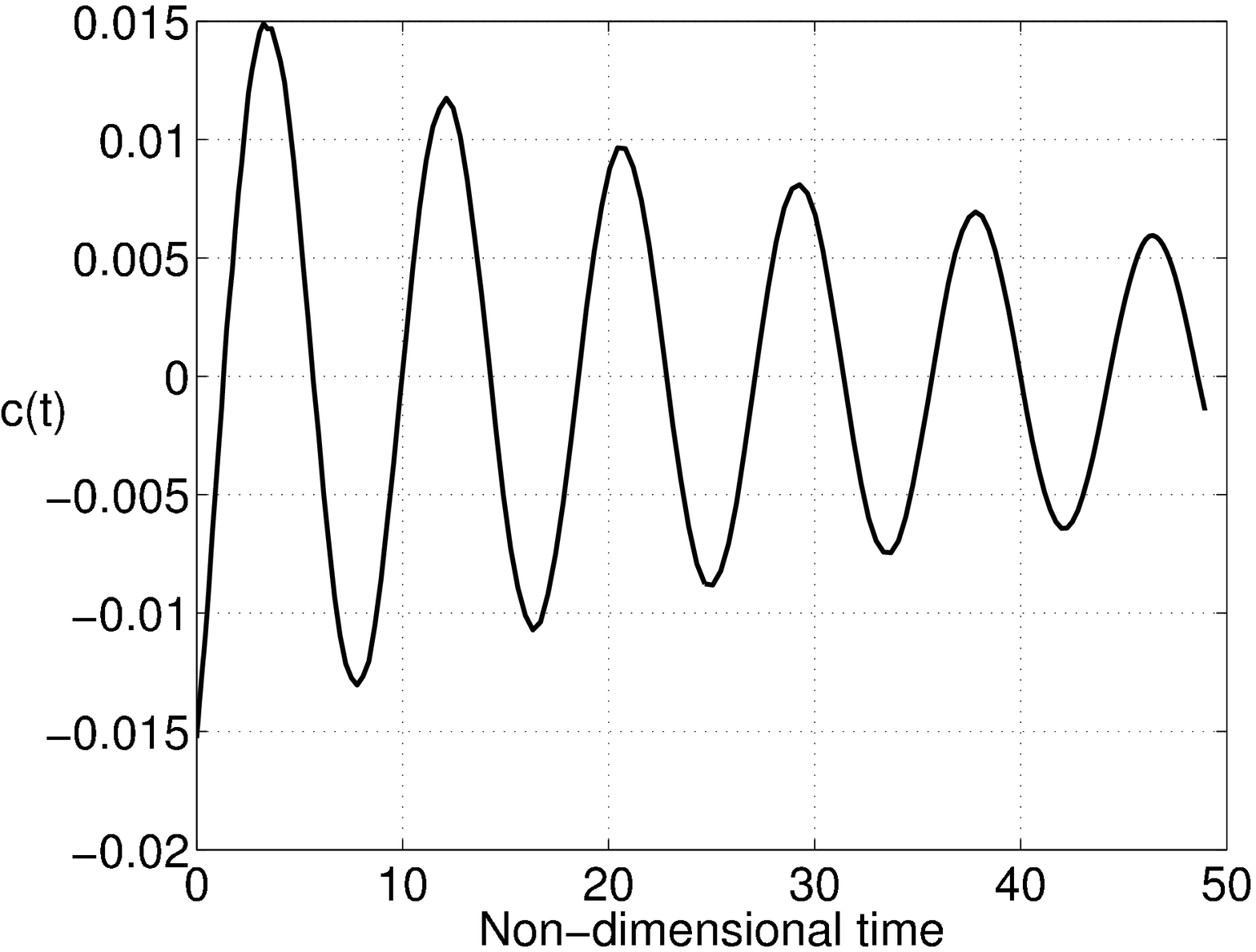}
  \includegraphics[width=3.5cm,height=3.5cm]{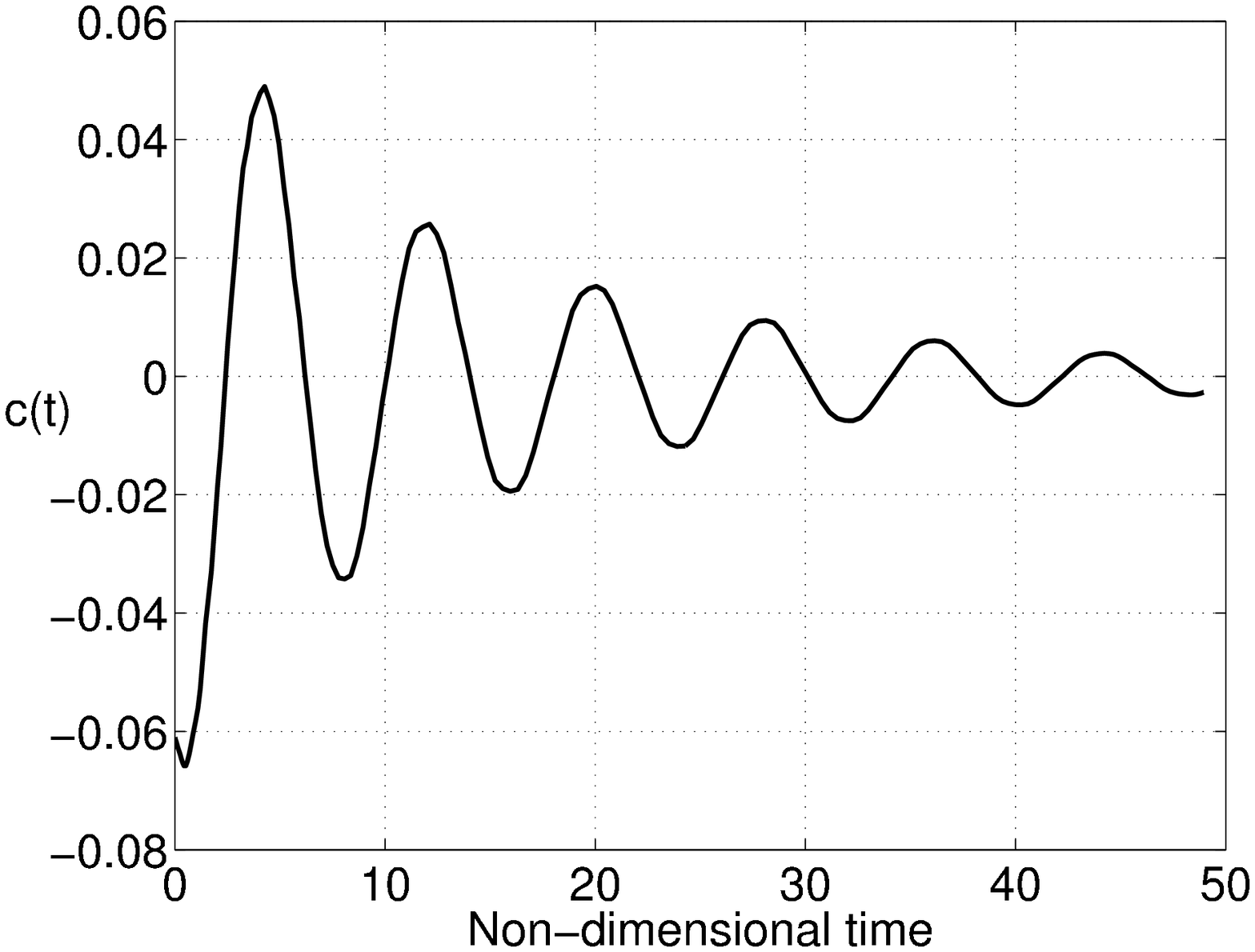}
  \includegraphics[width=3.5cm,height=3.5cm]{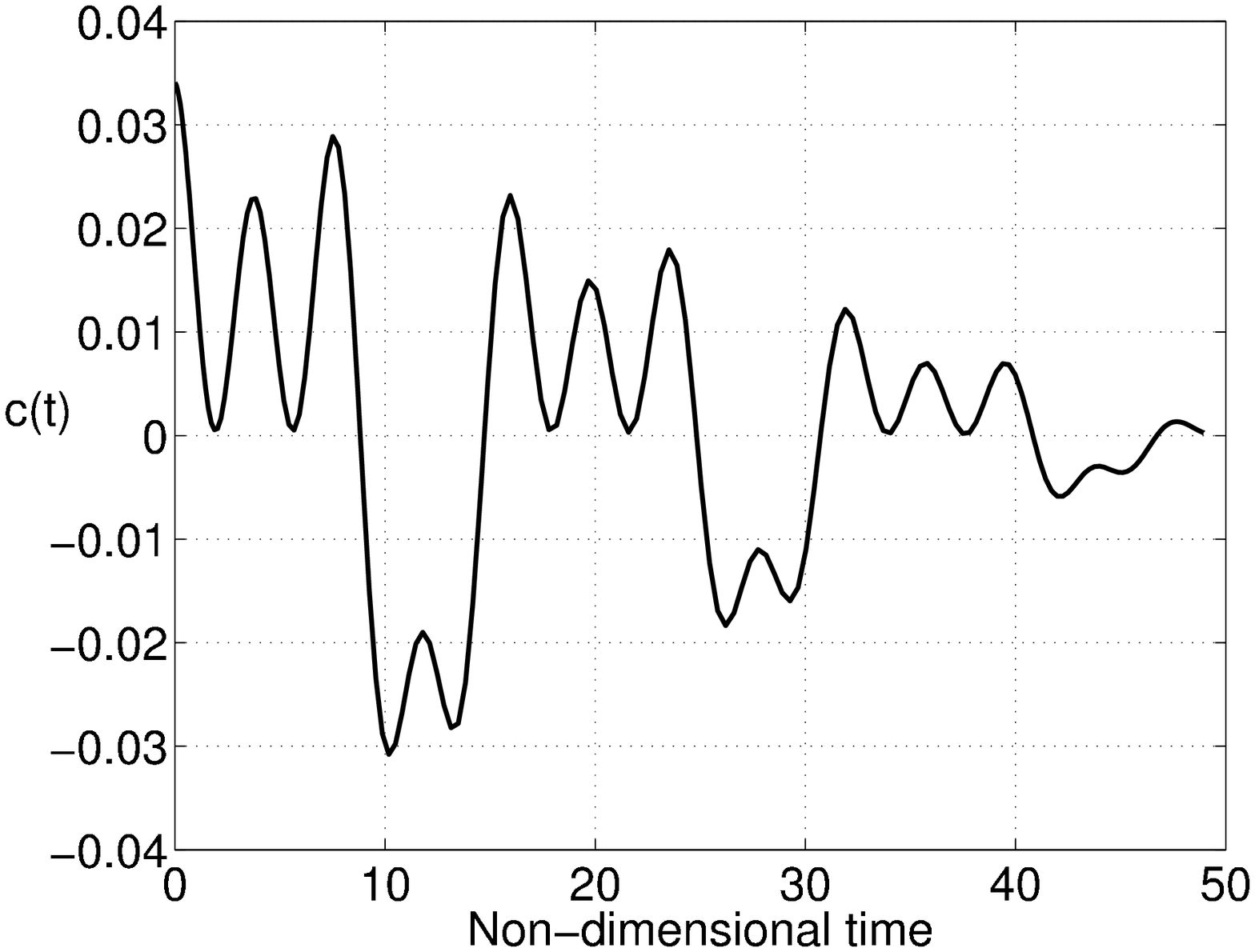}\\
\end{figure}
Eleven extra control laws were used for testing. A few examples are
plotted in Fig. \ref{fig:test60}. For these examples we also plot the
third modal coefficient obtained by projection and by model integration. 
Some discrepancies in coefficient amplitude are
observed, but overall the model predicts the right time dynamics.\\
In Fig. \ref{fig:multi60} we look at the results obtained with the
different models, using the distances and errors described
above. The first point to be made is that the
model error is almost zero when the distance from the model is
zero. This confirms that 1-control models work well when integrated
with the control law to which they were fitted. The errors then
increase with the distance from the model, as was expected. \\
The graph highlights the disadvantage of 1-control models. In the
best case the difference between projection and prediction
coefficients becomes higher than 20\% as soon as the distance from the
model exceeds 40\%. In contrast, for the 2-control and 3-control
models, the error stays under 20\%, even when the distance increases. 
\begin{figure}[]
 % \vspace{-4.8cm}  
  \centering 
  \includegraphics[clip,height=2.58cm]{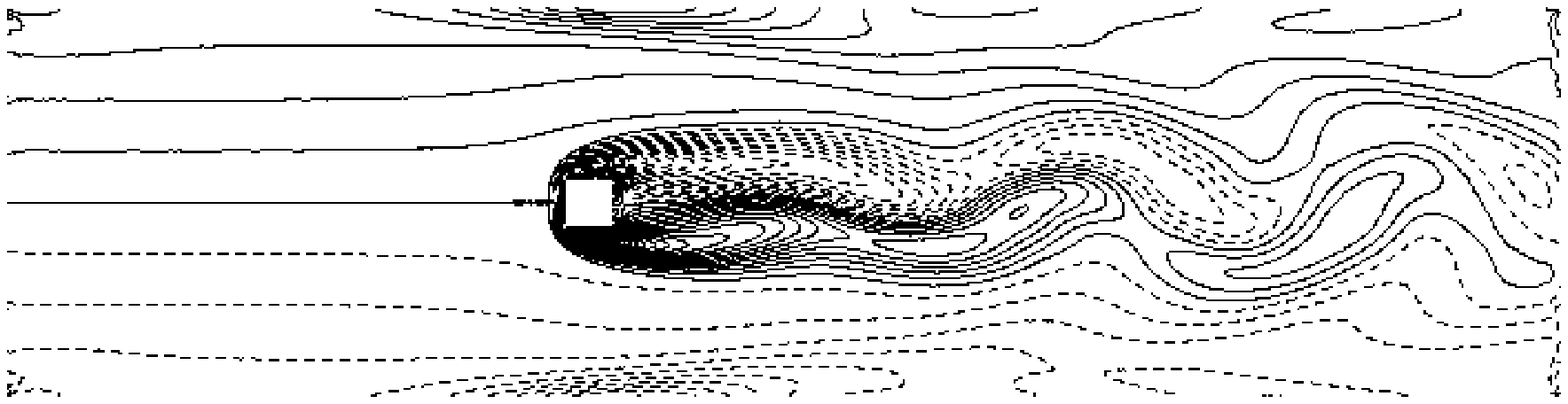}\\
 % \vspace{-9.cm}  
  \includegraphics[clip,height=2.5cm]{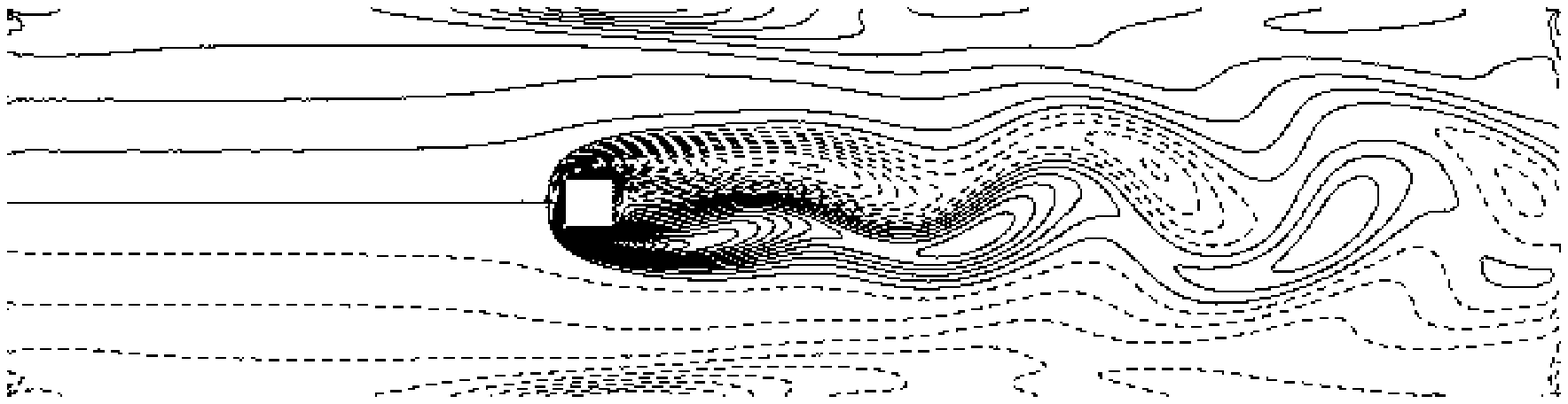}\\
 % \vspace{-4.cm}  
  \caption{Model predicted vorticity field (top) and Navier-Stokes
    vorticity field at $t=T$. Positive (continuous lines) and negative
    (dashed lines) vorticity isolines} 
  \label{fig:ric60}
\end{figure}
In Fig. \ref{fig:ric60} we plot isolines of the vorticity at time
$t=T$ for one of the test control laws (the third control law in
Fig.\ref{fig:test60}). Time coefficients were obtained by
solving $\M{R}({\M{C}})$ with ${\M{C}}=\{c_1,c_2\}$. The velocity field was then
reconstructed using the first ten of these coefficients and the first
ten POD modes in $\phib({\M{C}})$. The reconstructed vorticity is
presented along with the vorticity obtained by running the Navier-Stokes
equations with the test control law. The controls used to build the
model caused a slight decrease in vortex size (see
Fig. \ref{fig:cal60}, bottom left) whereas actuation used
in the test caused a slight increase in vortex size (see
Fig. \ref{fig:test60}, bottom right) . We note that the
model was able to predict such features, and that at the end of the
simulation time, the structure of reconstructed flow is almost
identical to that of the real flow. In contrast, the 1-controls
weren't able to identify this. If the same reconstruction is performed
using ${\M{C}}=\{c_1\}$ for example, the flow appears almost stable at
$t=T$, meaning the model predicted the opposite behavior to what
actually happened.  

\subsection{Results for $Re = 150$}
\label{sub:re150}

\begin{figure}[!t]
  \centering 
  \includegraphics[width=3.95cm,height=6.5cm]{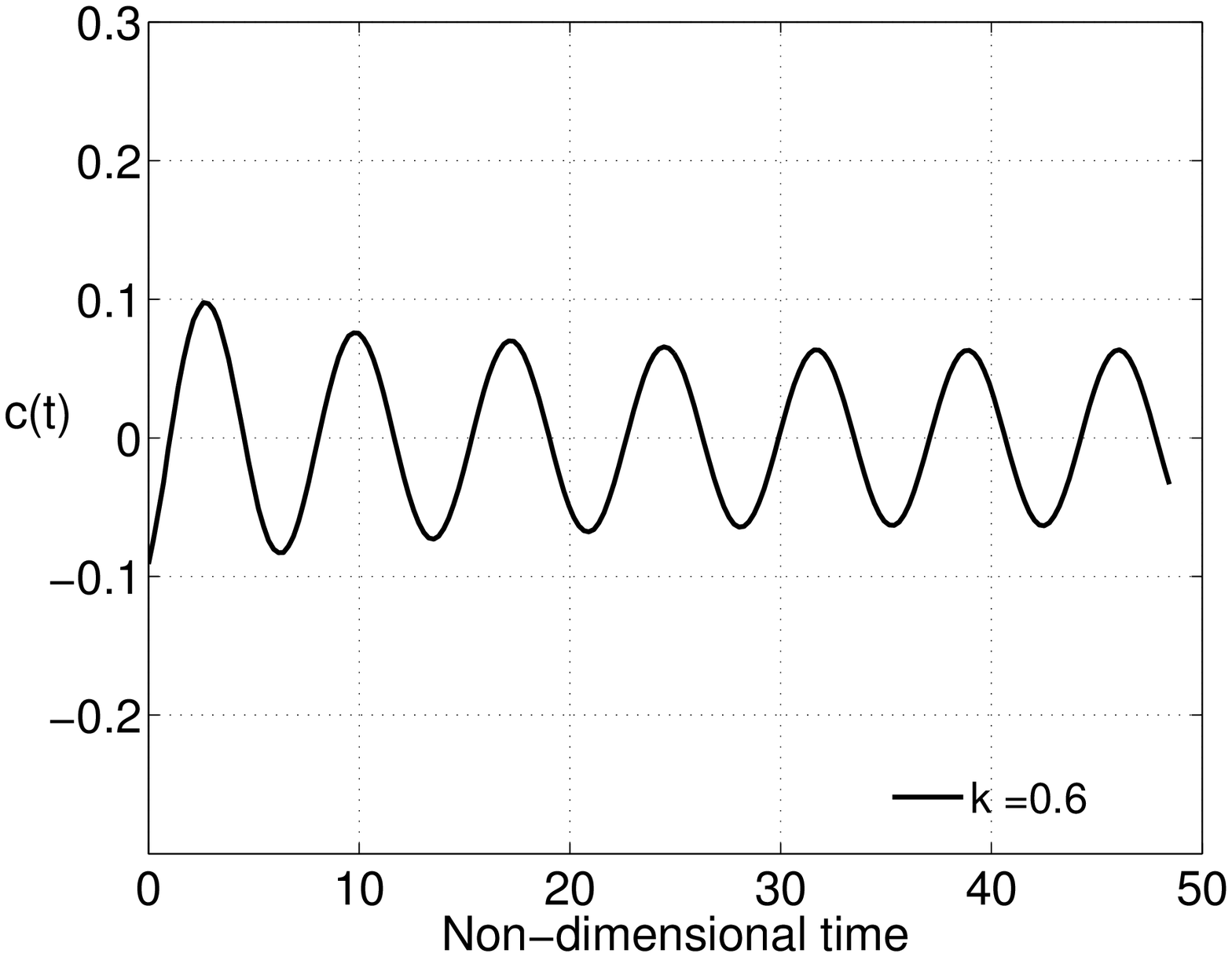}
  \hspace{-0.5cm}
  \includegraphics[width=3.95cm,height=6.5cm]{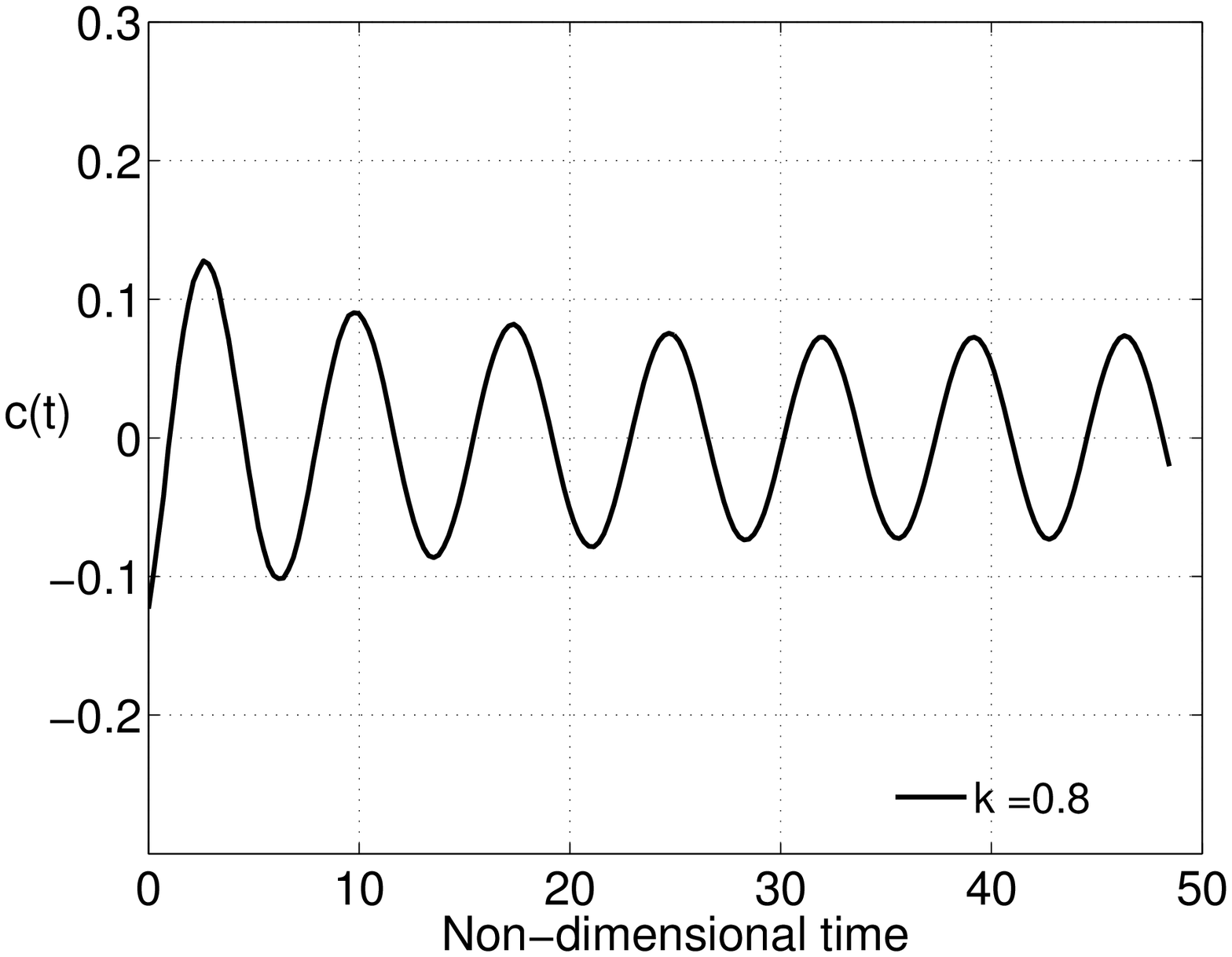}
  \hspace{-0.5cm}
  \includegraphics[width=3.95cm,height=6.5cm]{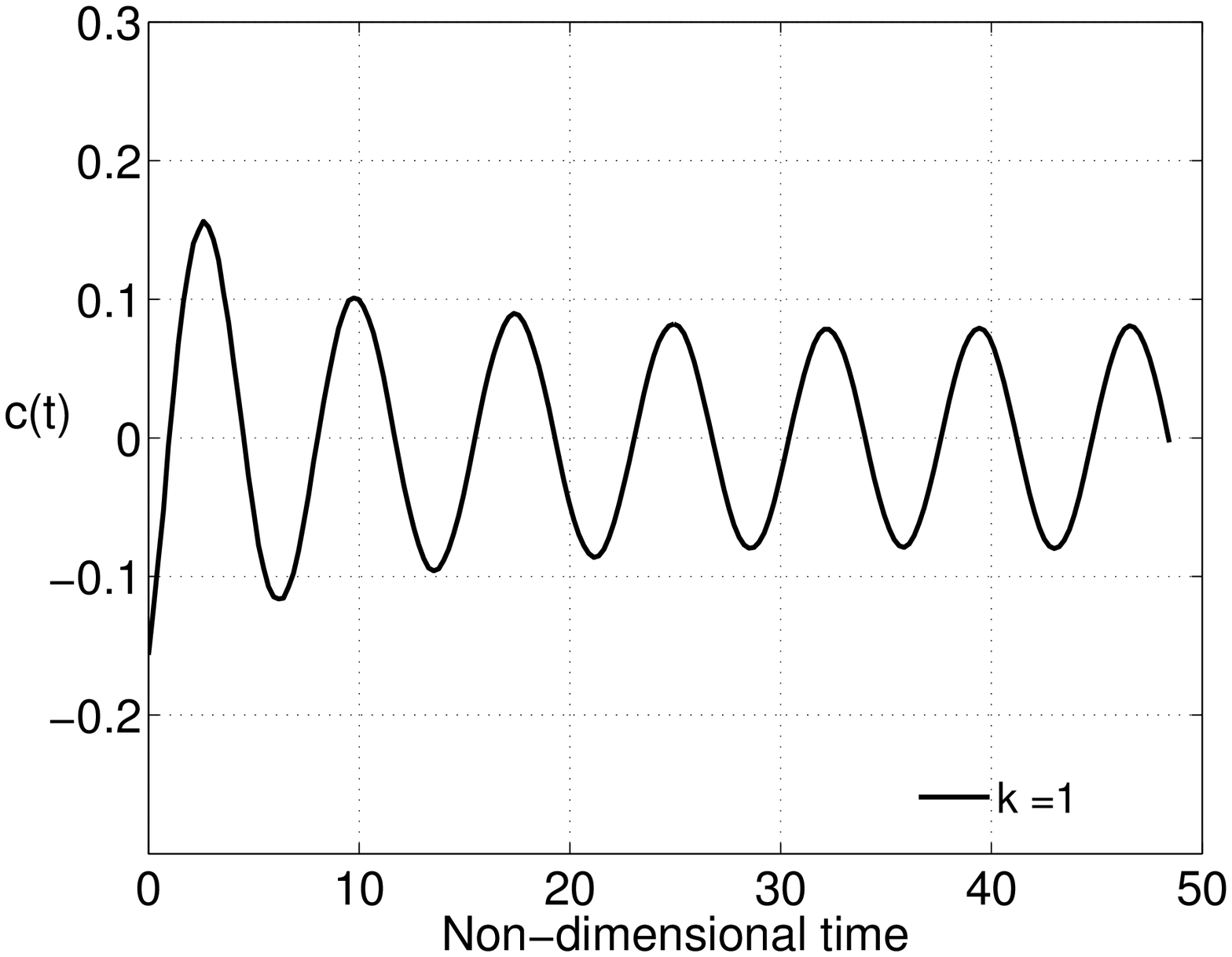}\\
  \vspace{-2.8cm}
  \hspace{0.1cm}
  \includegraphics[width=3.95cm,height=6.5cm]{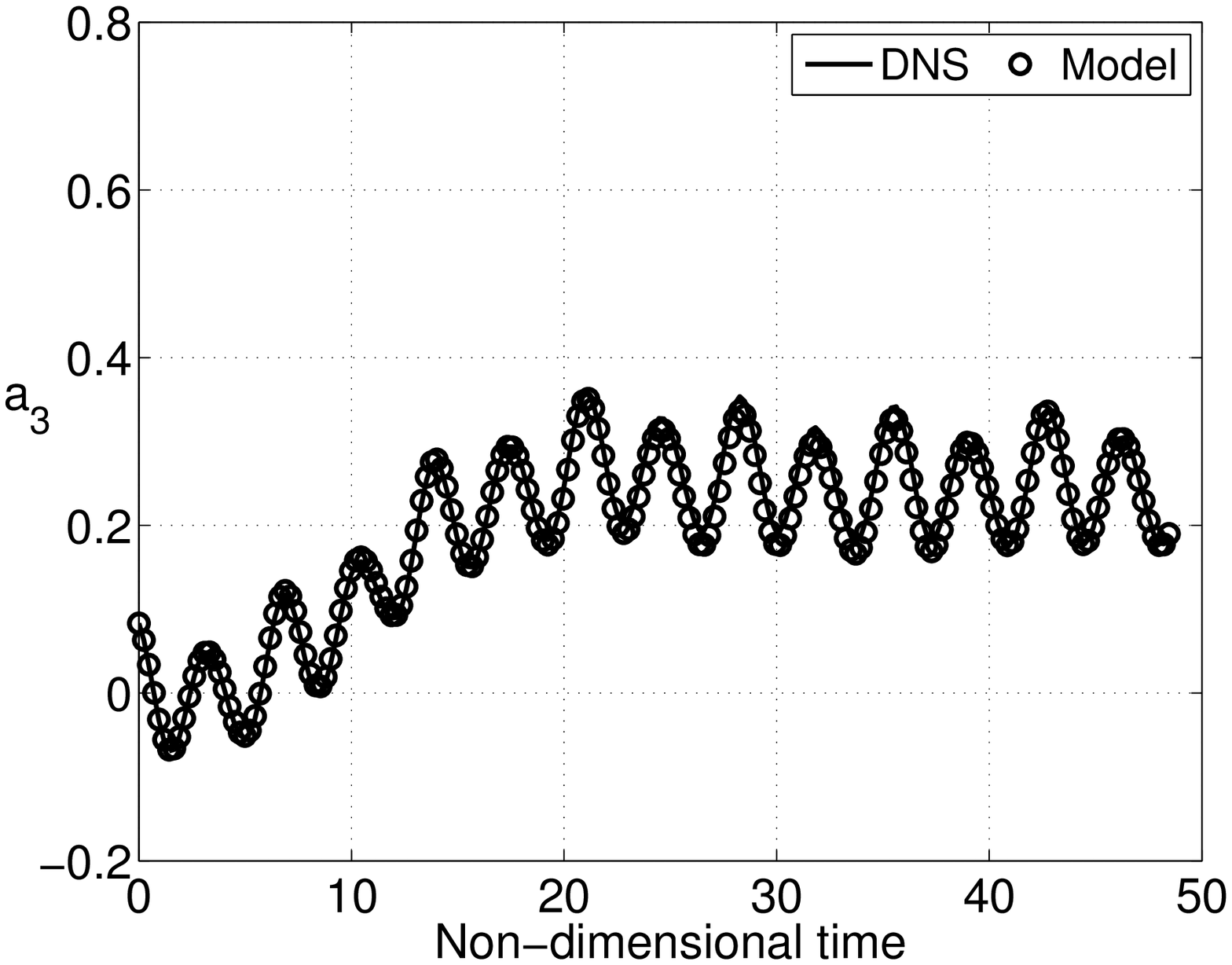}
  \hspace{-0.5cm}
  \includegraphics[width=3.95cm,height=6.5cm]{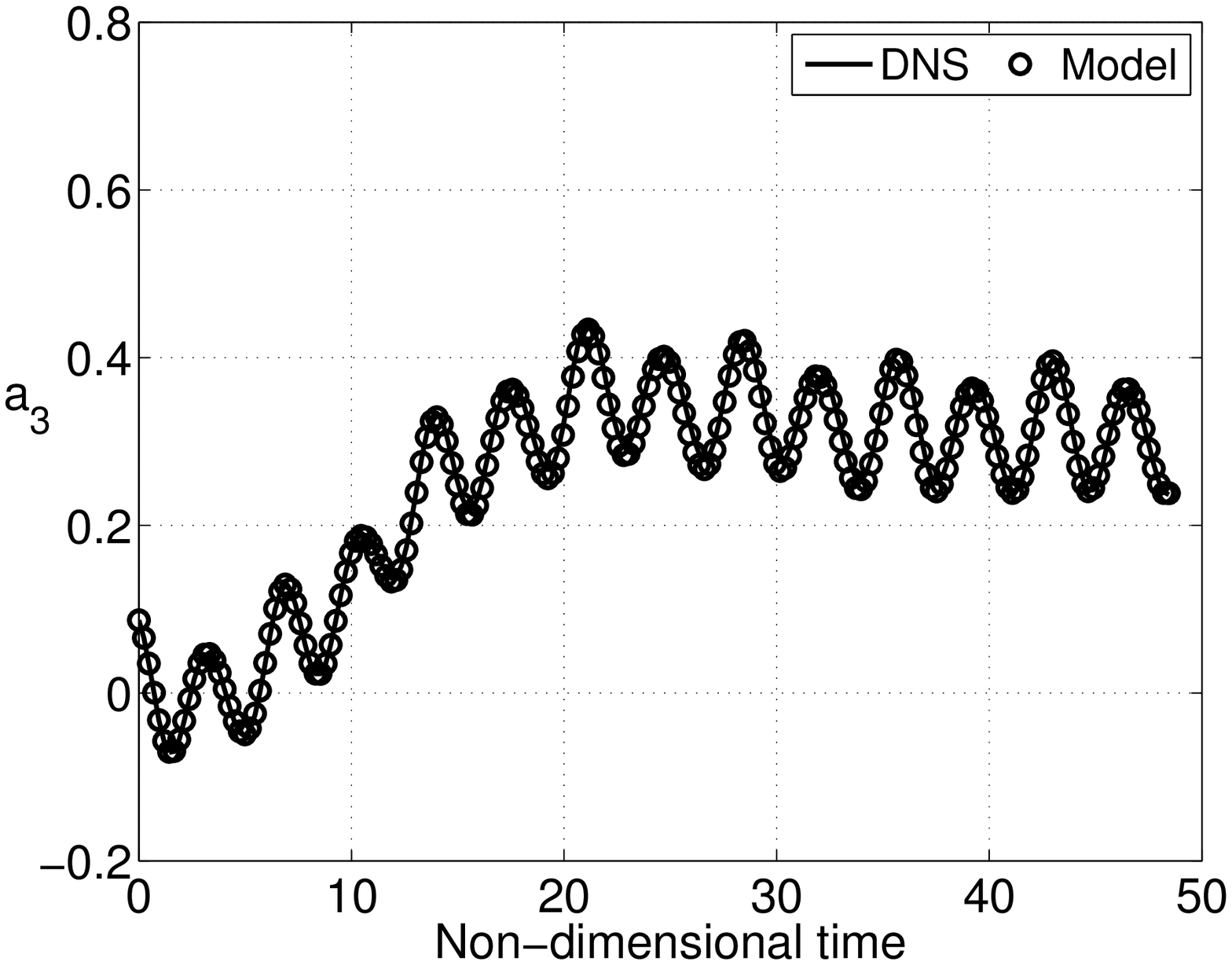}
  \hspace{-0.5cm}
  \includegraphics[width=3.95cm,height=6.5cm]{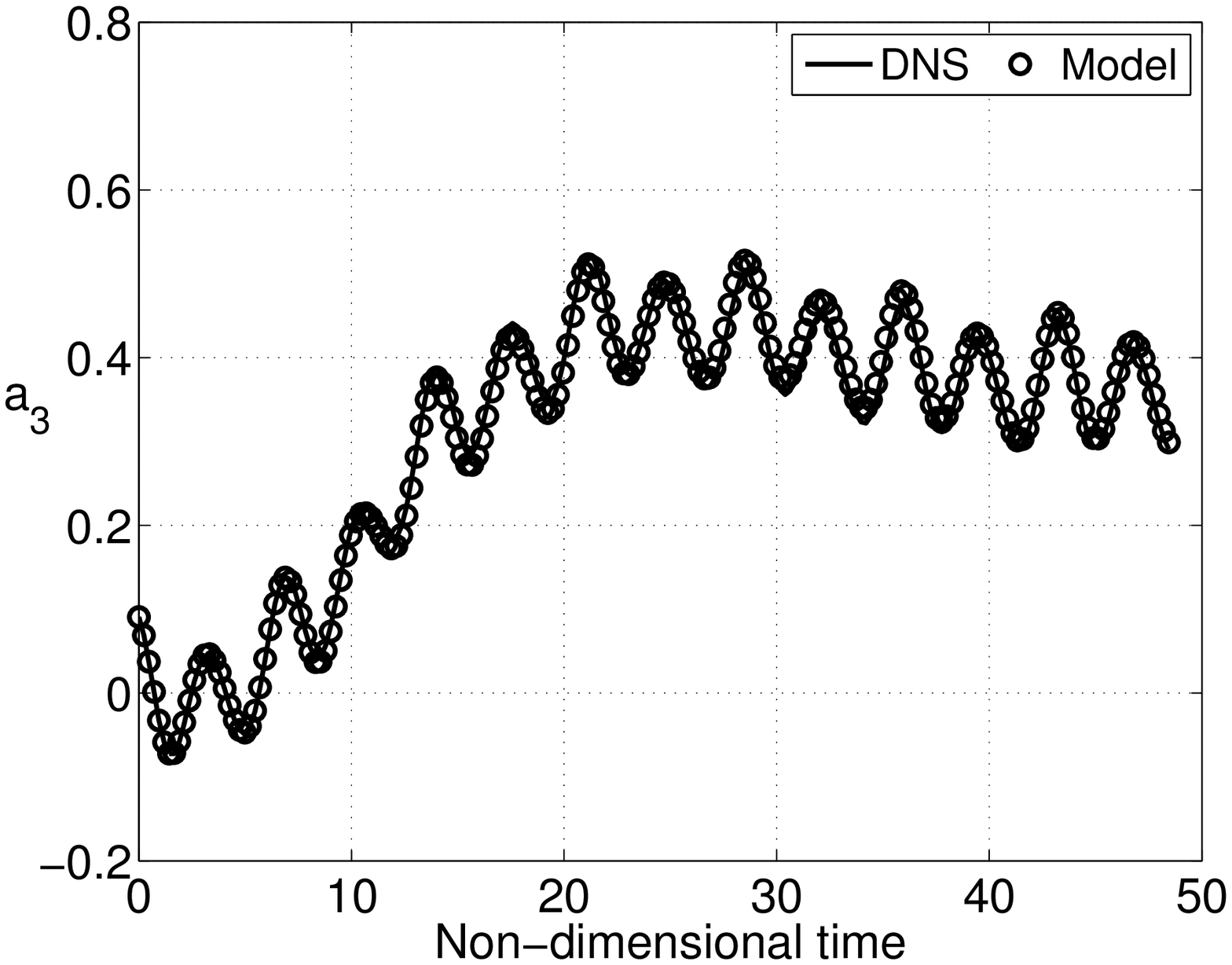}  
  \vspace{-1.5cm}  
  \caption{Control laws used to build the models (top); $a_3$ DNS
    (continuous line) vs prediction by 3-control model (symbols)}
  \label{fig:cal150}
\end{figure}

For $Re = 150$ only feedback control laws are used both to build
the models and to perform the tests. In Fig.\ref{fig:cal150} the three
feedback control laws used to calibrate the model are shown. The laws
are obtained with sensors placed at $(x_s,y_s)=(0.7,0.0)$ and by using
gains $\Kb=0.6$, $\Kb=0.8$ and $\Kb=1$. The figure also
shows the third modal coefficients given by integrating the 3-control
feedback model with each gain. Although the control laws induce
different dynamics, the model is able to give an accurate prediction
in all three cases. 
\begin{figure}[!t]
  \centering 
  \includegraphics[width=3.95cm,height=6.5cm]{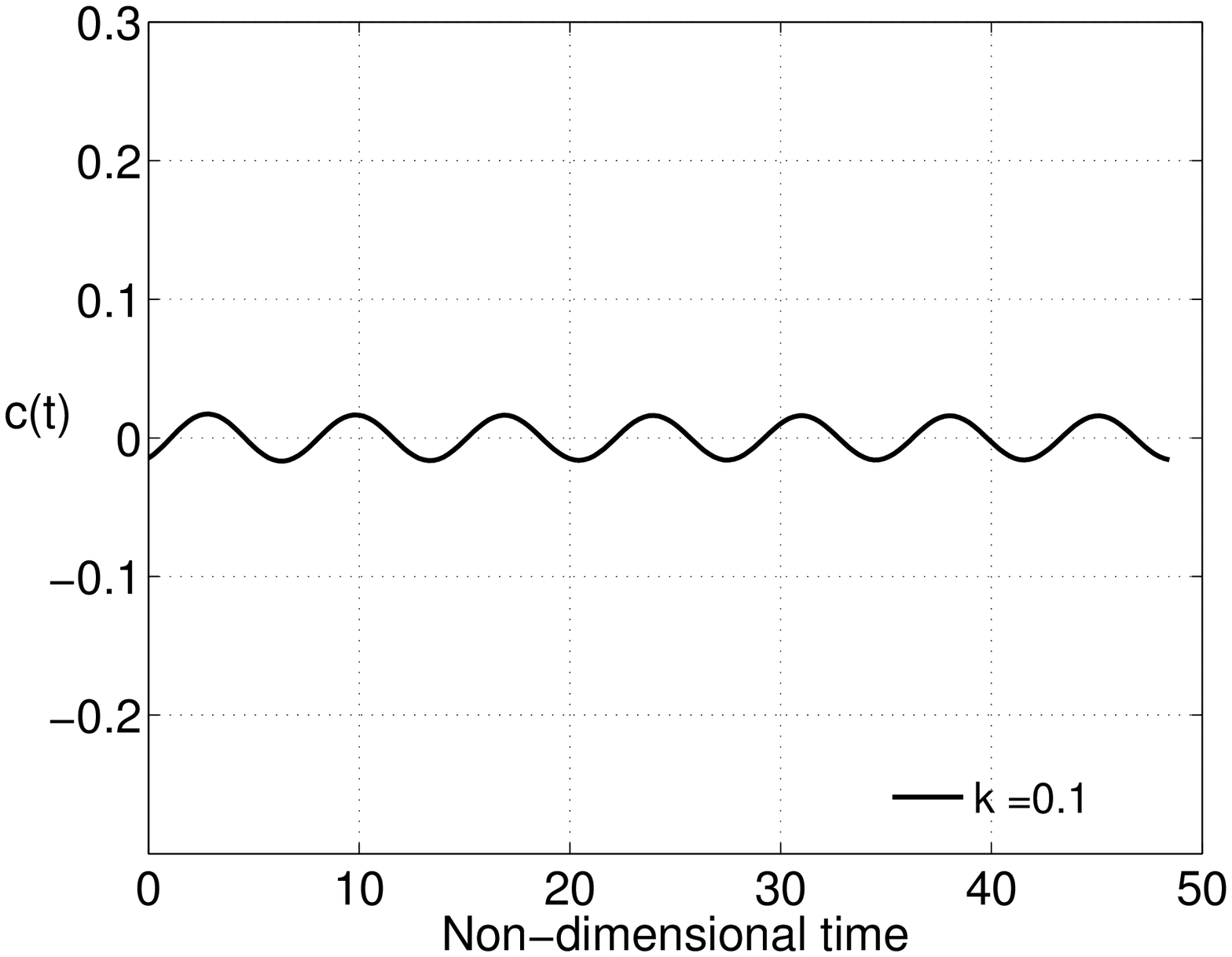}
  \hspace{-0.5cm}
  \includegraphics[width=3.95cm,height=6.5cm]{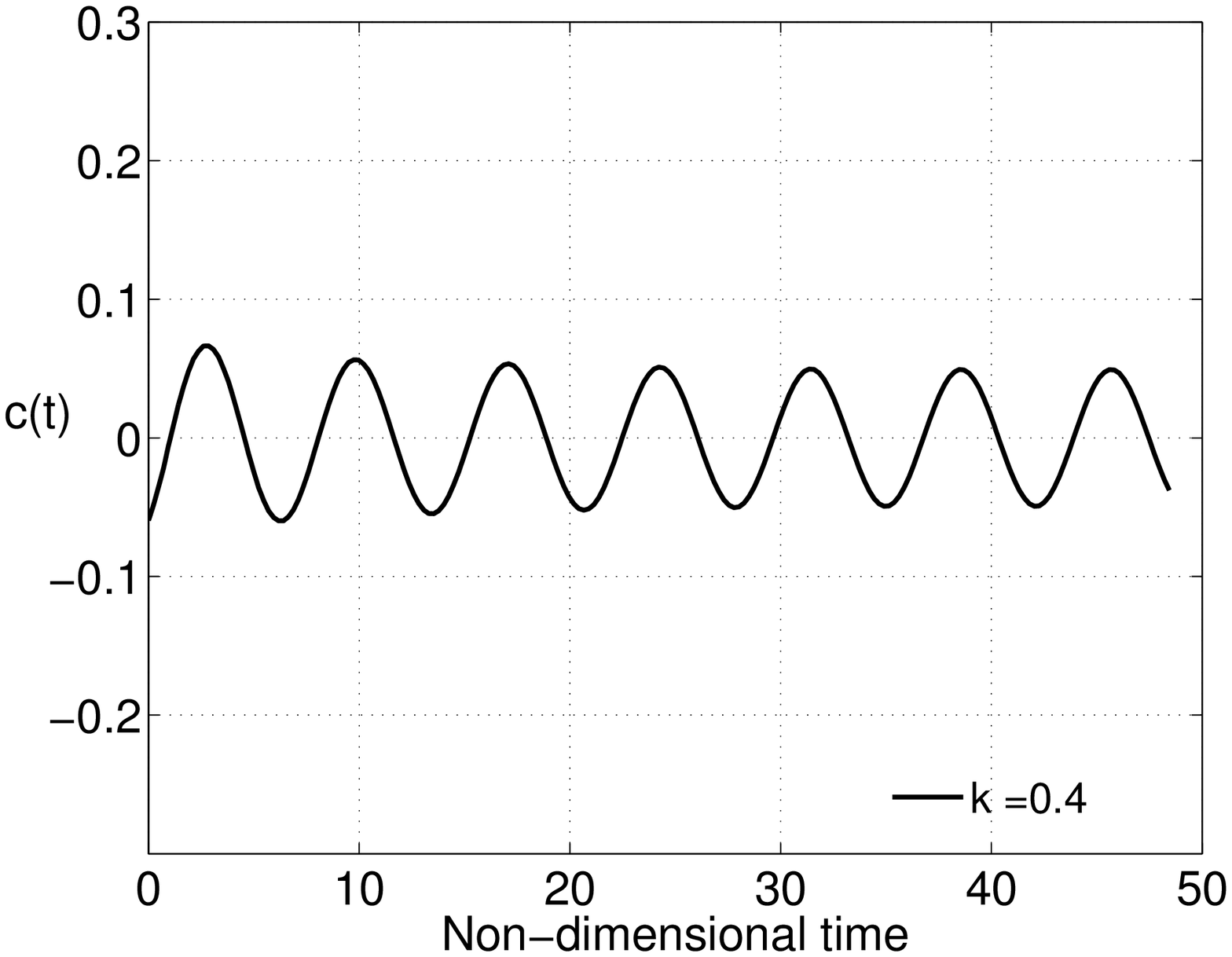} 
  \hspace{-0.5cm}
  \includegraphics[width=3.95cm,height=6.5cm]{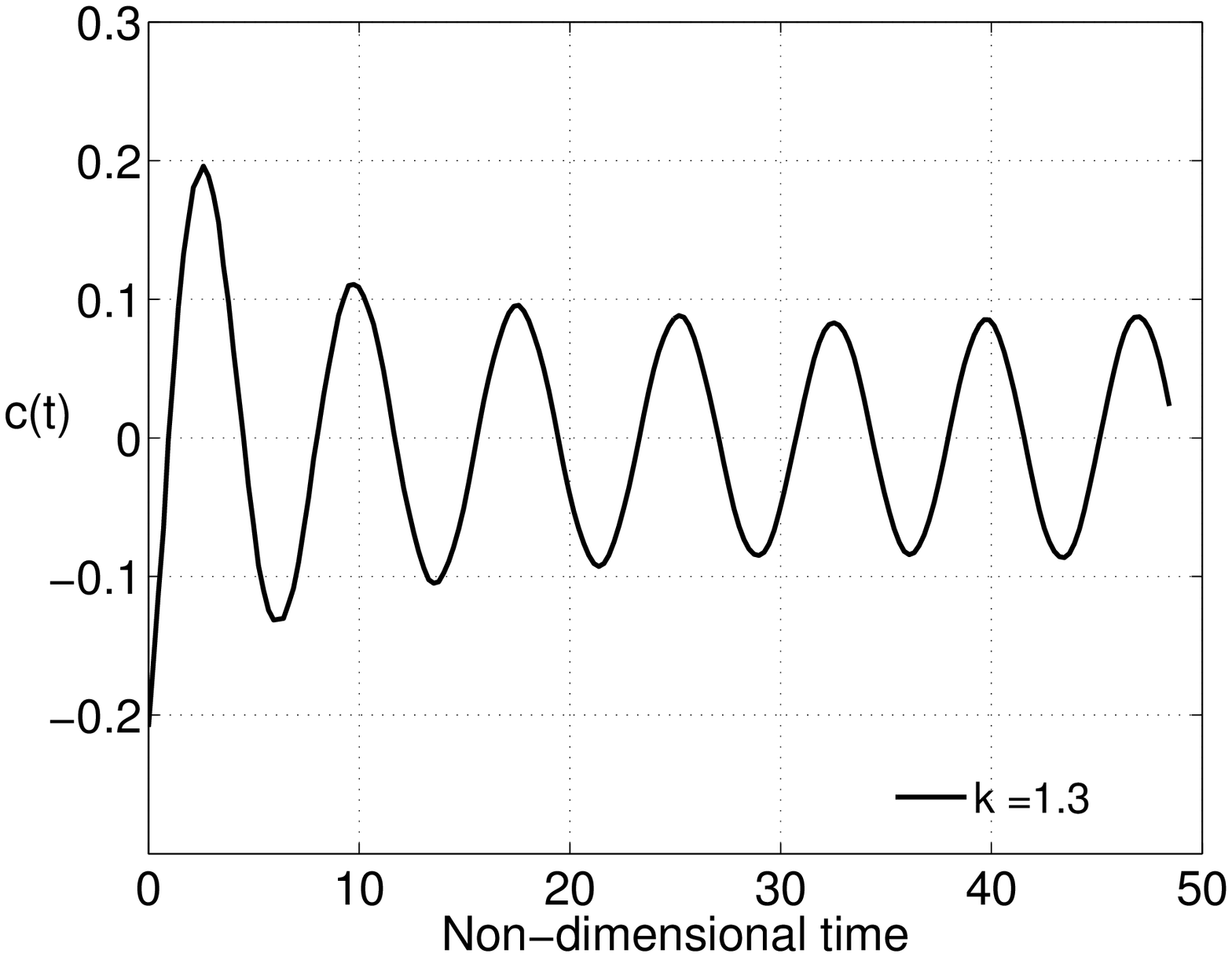} \\
  \vspace{-2.8cm}
  \hspace{0.1cm}
  \includegraphics[width=3.95cm,height=6.5cm]{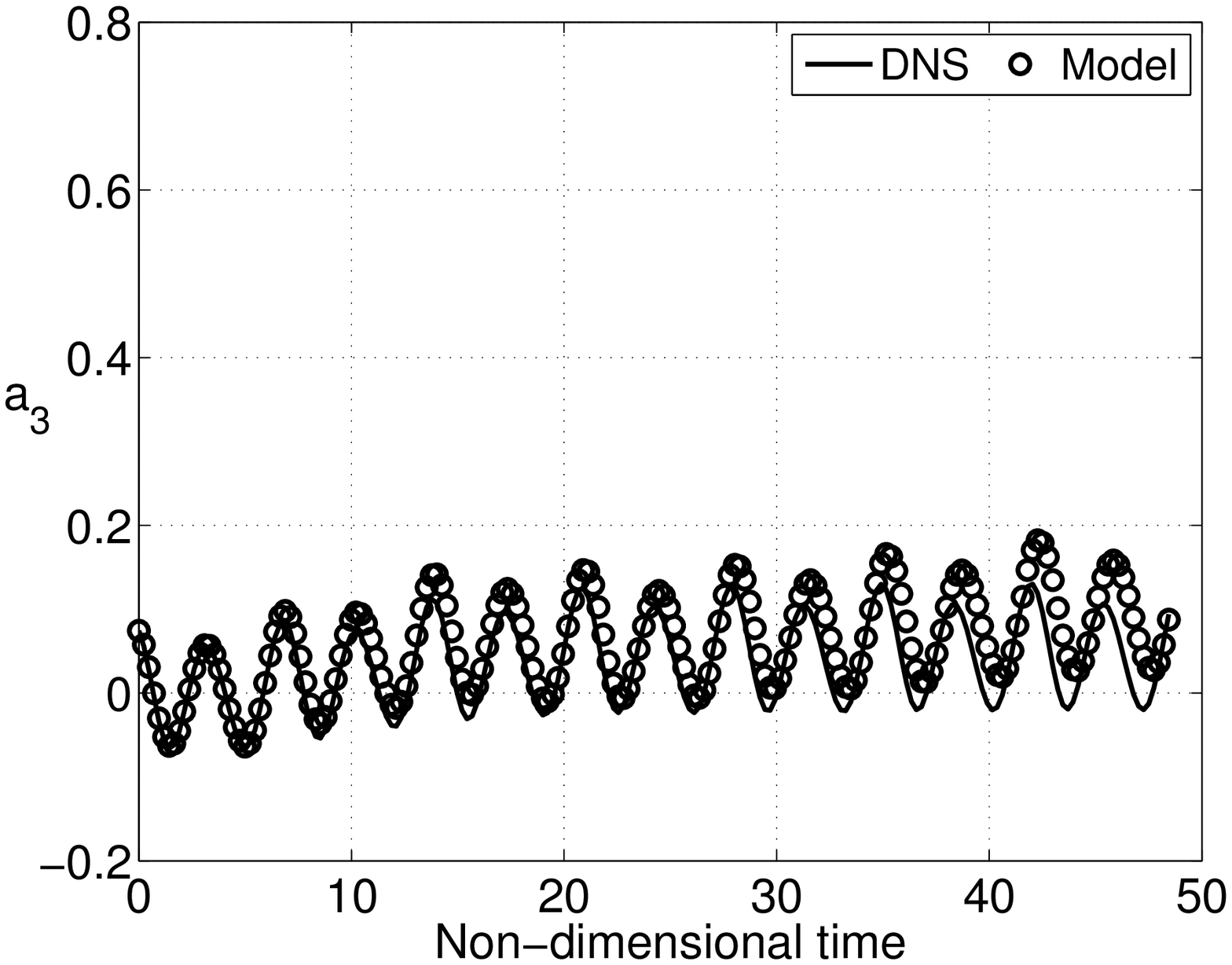}
  \hspace{-0.5cm}
  \includegraphics[width=3.95cm,height=6.5cm]{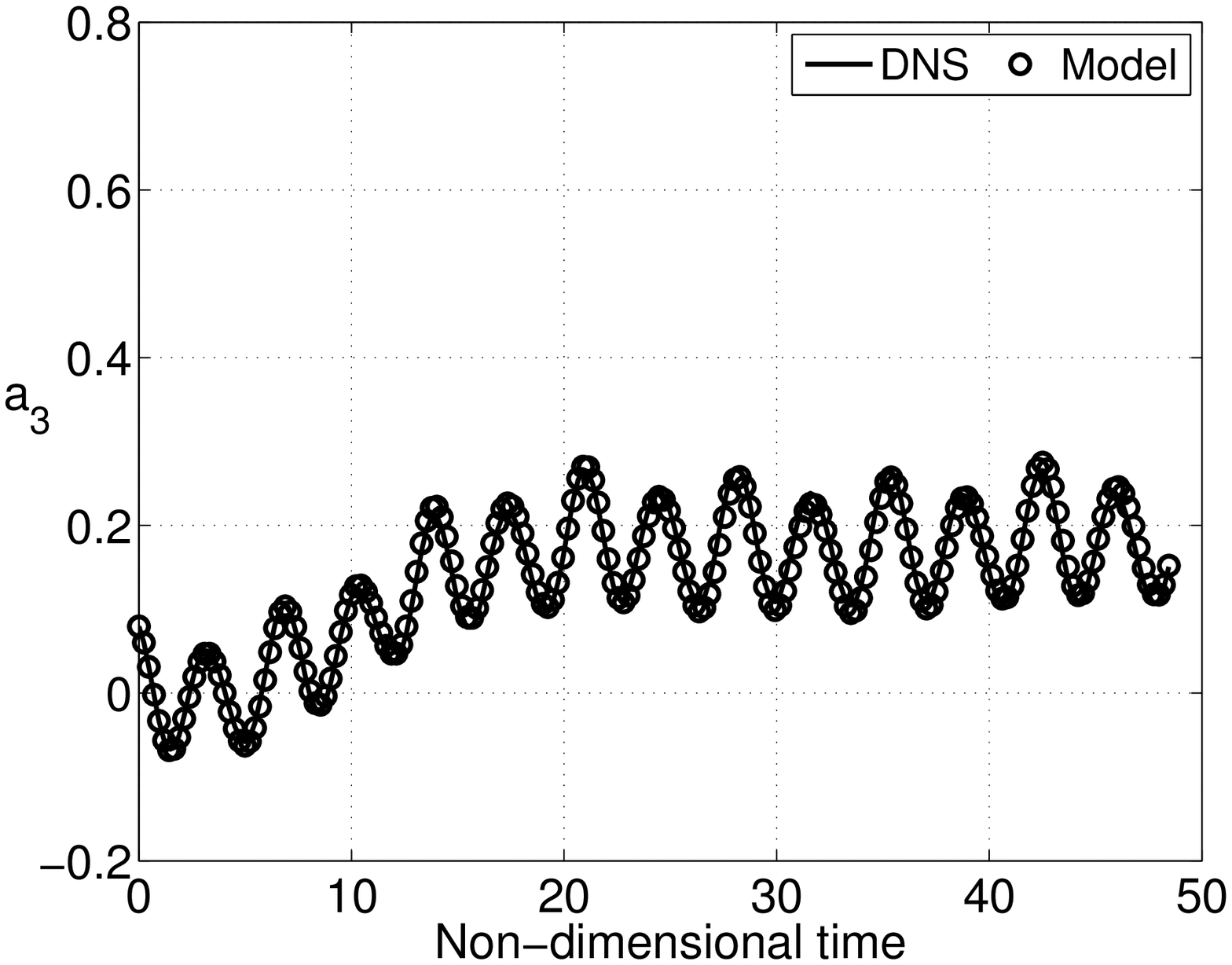}
  \hspace{-0.5cm}
  \includegraphics[width=3.95cm,height=6.5cm]{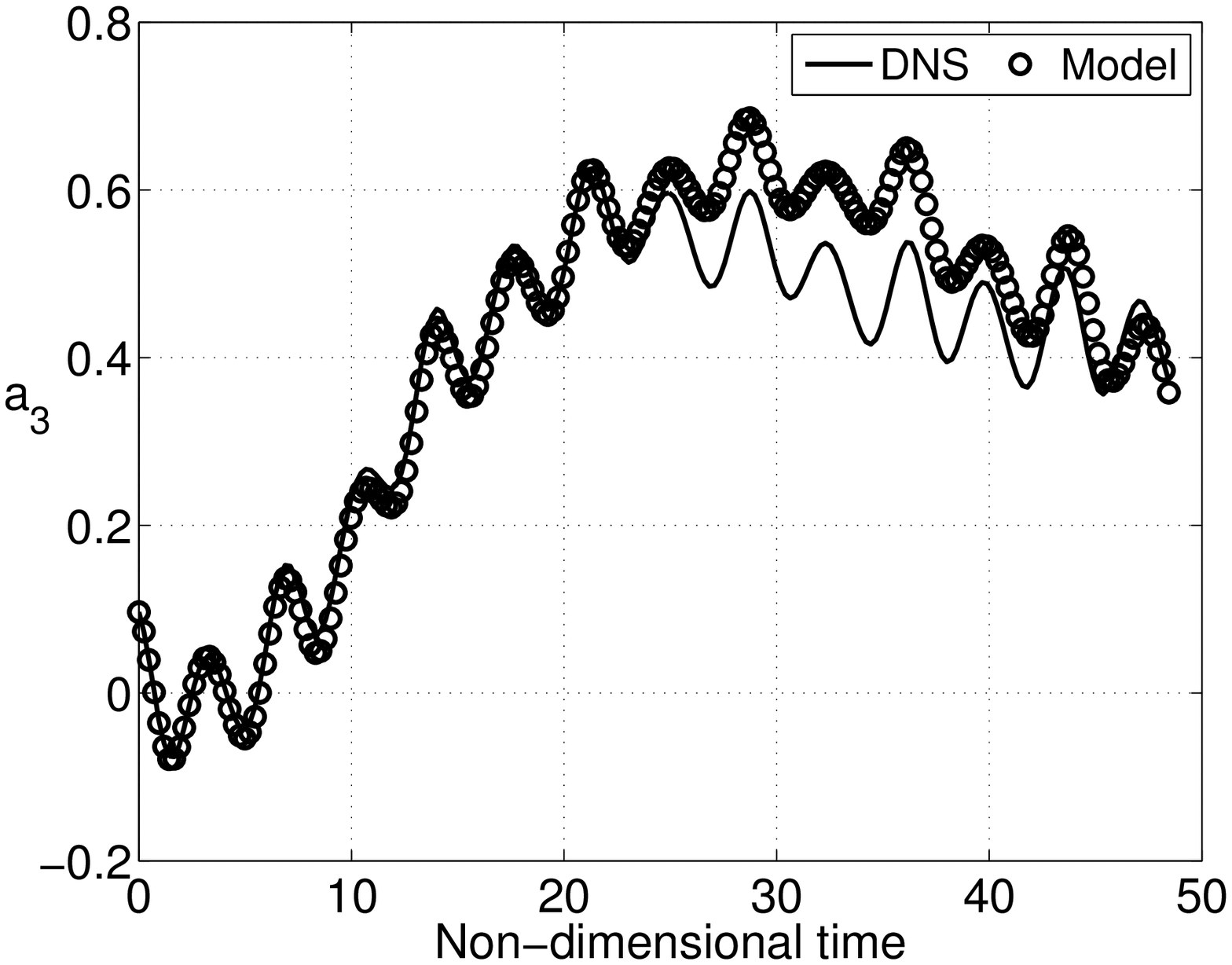}    
  \vspace{-1.5cm}  
  \caption{Control laws used to test the models (top); $a_3$ DNS
    (continuous line) vs prediction by 3-control model (symbols)}
  \label{fig:test150}
\end{figure}
Six extra control laws were used for testing, each corresponding to a
different choice of $\Kb$. A few examples, with corresponding
coefficients $\ah_3(t)$ are plotted in Fig.\ref{fig:test150}. It
appears that the dynamics are quite different when the distance,
between the gain value and gains included in the model, is large. For
example, when using a gain $\Kb=0.1$ the average value of $\ah_3(t)$
is low compared to that obtained with $\Kb=1$. However, the 3-control
model again gives an overall good prediction of the time dynamics. 
\begin{figure}[!t]
  \centering 
  \includegraphics[clip,width=8.5cm,height=12.5cm]{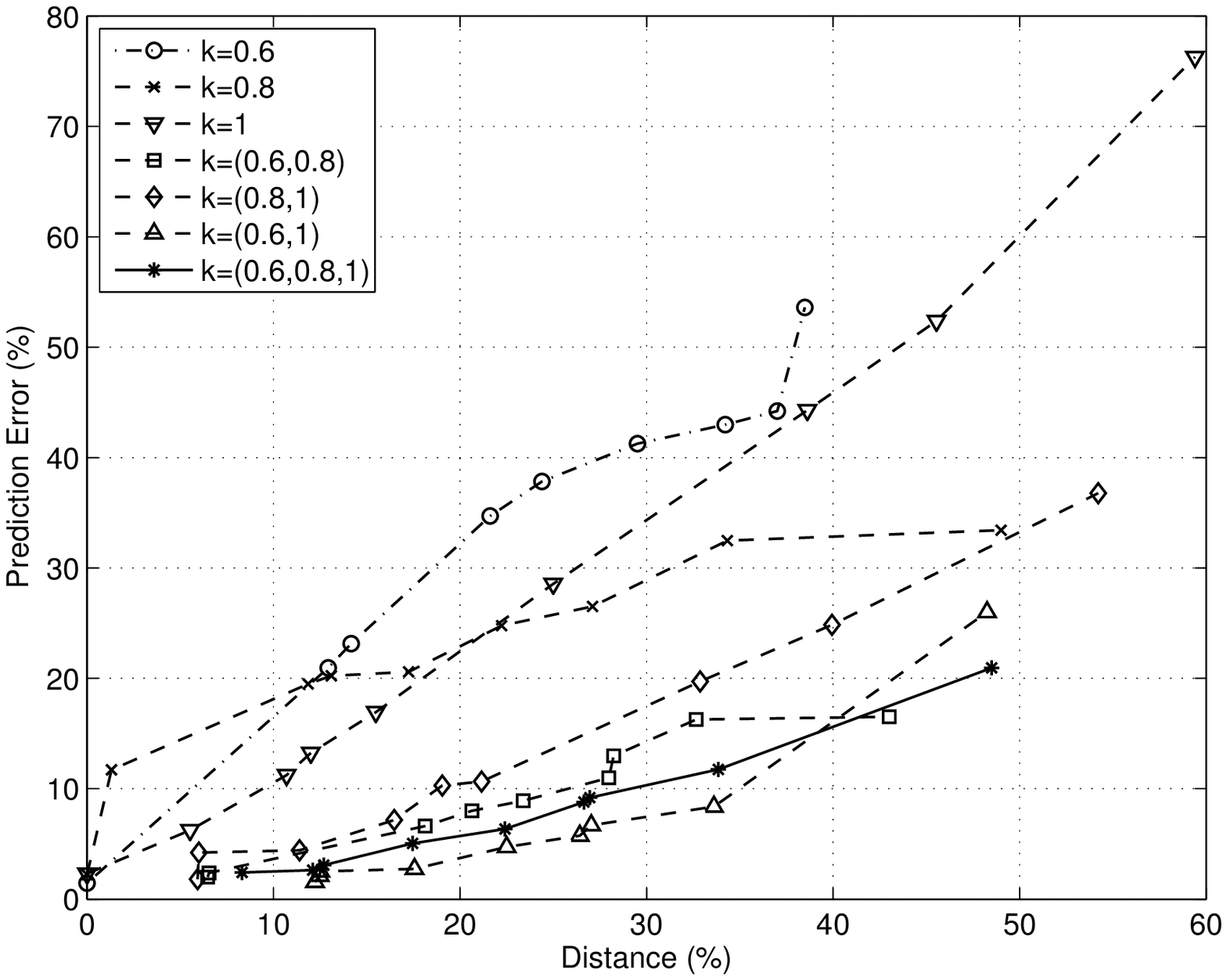}
  \vspace{-3.cm}
  \caption{Prediction errors obtained using 1-control, 2-control and
    3-control models}
  \label{fig:multi150}
\end{figure}
\begin{figure}[!t] 
%  \vspace{-3.5cm}
  \centering
  \includegraphics[height = 2.55cm]{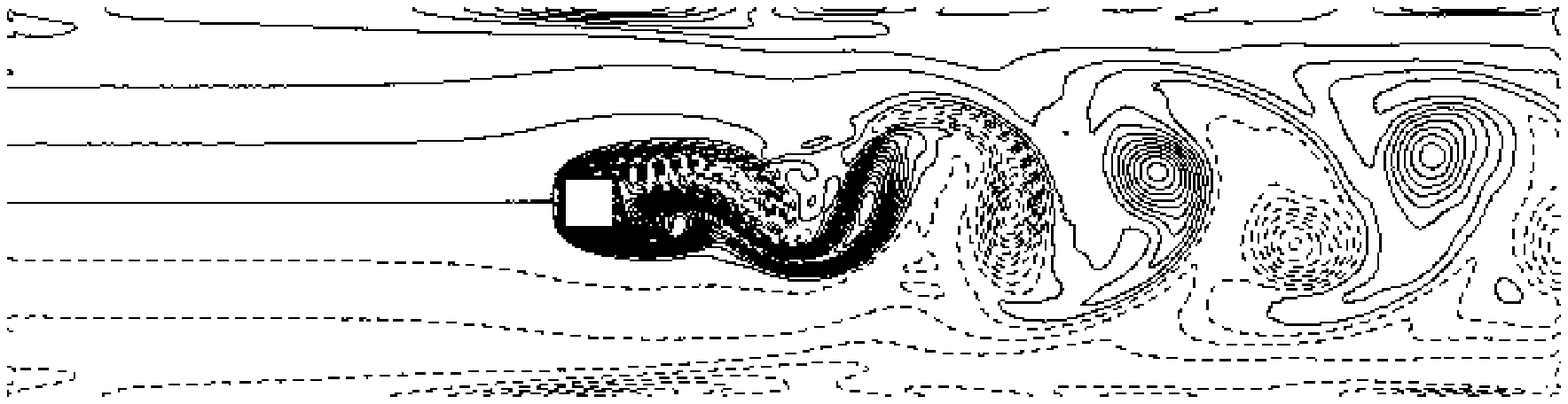}\\
%  \vspace{-9.cm} 
  \includegraphics[height = 2.5cm]{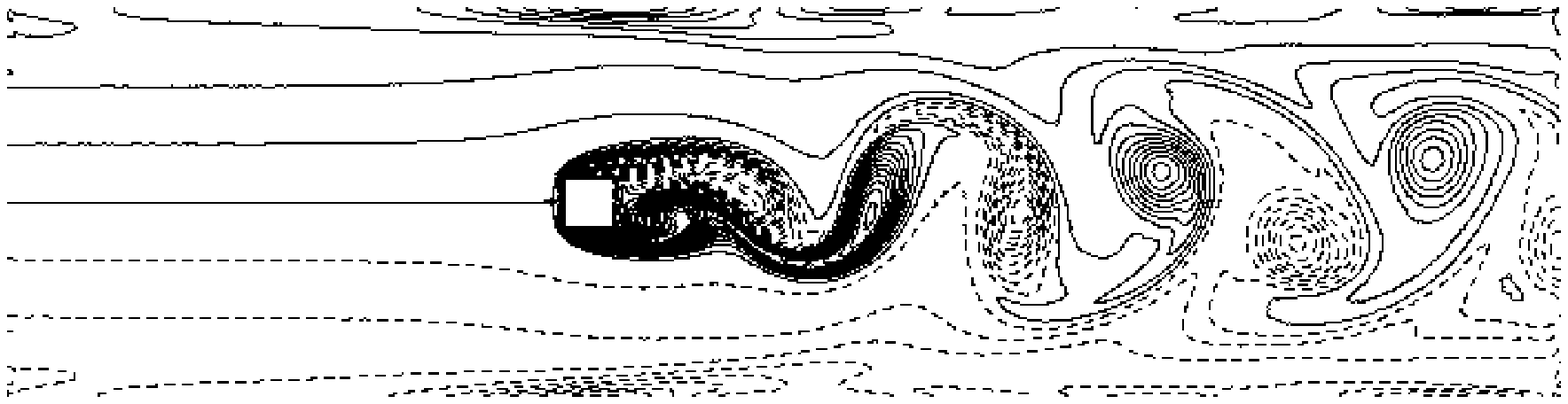}\\
%  \vspace{-4.cm}
  \caption{Model predicted vorticity field (top) and Navier-Stokes
    vorticity field at $t=T$ obtained with $\Kb=0.1$. Positive
    (continuous lines) and negative (dashed lines) vorticity isolines}
  \label{fig:ric150}
\end{figure}
Fig.\ref{fig:multi150} is built the same way as
Fig. \ref{fig:multi60}. In particular the graph shows the disadvantage
of using a 1-control model, with prediction errors of over $34\%$ when
the distance from the calibration dynamics increases over the
$30\%$. As in the case $Re=60$, the 2-control models give more
accurate predictions than the 1-control models. The lowest errors are
obtained with the 2-control model $(\Kb=0.6,\Kb=0.1)$. This
observation suggests that, in model construction, an optimized a
priori choice of the sampling points could be useful to obtain a more
robust model. We note that in this case it was the model built to fit
the highest and lowest values of $\Kb$ that gave the best result, and
that adding a third intermediate control to the model ($\Kb=0.8$) did
not bring any improvement: the 3-control model gives more or less the
same results. \\  
In Fig. \ref{fig:ric150} we plot isolines of the vorticity at time
$t=T$ for the flow obtained using $\Kb=0.1$ as feedback gain (the
first one in Fig.\ref{fig:test150}). Time coefficients were obtained by
integrating the 3-control model. The velocity field was then reconstructed using all
the $60$ coefficients and POD modes. The reconstructed vorticity is
presented along with the vorticity obtained by running the Navier-Stokes
equations with the test control law. The controls used to build the
model were similar in the sense that they had a much stronger effect on the
flow compared to the control obtained with $\Kb=0.1$. We note that the
model is able to accurately predict a flow snapshot and that the
reconstructed flow is almost identical to that of the real flow.

\section{Conclusions}

The overall picture of reduced-order modeling that results from our
study is the following. Given a control law, one can deduce a
low-order model of the actuated flow by simply projecting the
Navier-Stokes equations on POD modes. The coefficients of the
quadratic model thus obtained are found by projection. However, a model
constructed this way will show large time-integration errors even for
the same control law used to generate  the POD modes. Calibration can take care of
that, in the sense that the model coefficients can be determined in
order to match as closely as possible at least the solution from which
the POD modes are obtained. This might lead to a numerically stable
model. However, this model is generally not at all robust, in the sense
that the predictions for a slightly different configuration from that
it was generated from, fails. A symptom of
such lack of robustness is observed in the ill-posedness of the
inverse problem: the matrices to be inverted are almost singular.\\ 
In order to get around this deficiency, we regularize the solution by
adding a constraint to the minimization method used to solve the
inverse problem. We ask that the coefficients of the polynomial
expansion be close enough to those obtained by projection. This
method allows to synthesize models that adequately simulate the flow in
a small vicinity of the control law used to generate the solution
database. However, the actual real improvement in robustness is
obtained by spanning the solution manifold, i.e., by including several
control laws in the inverse problem definition. By doing this, the
results presented show that the models are able to predict dynamical
behaviors that are far, in terms of an energy norm, from the cases
included in the database. A consequence of such an additional
regularization is that the matrices involved in the inverse problem
solution become well conditioned. \\
Another important aspect of the
method proposed, is that its cost is that of a matrix inversion, and
that it does not scale with the number or the size of data sets used to build
the model. Therefore it seems reasonable to envisage an automatic
strategy to enrich the model by spanning the control space. In this
respect, the technique proposed in~\cite{Willcox2007} to distribute
in an optimal way the points where to test the control space can help
minimize the number of a priori simulations needed to build the
model. For example, our results show that a model based on two
controls might predict the effect of actuation laws not present in the
data base, as precisely as a model based on three controls, if the two
controls are appropriately placed. \\
In conclusion, the modeling we propose appears to be a viable approach
to determine control strategies for those problems that because of their
computational size cannot be treated in the framework of classical
control theory. 

\bibliography{bib_rid}
\bibliographystyle{plain}

\end{document}